\documentclass[preprint]{aastex631}

\usepackage{xcolor}
\usepackage{subfloat}
\usepackage{multirow}
\usepackage{soul}

\usepackage{graphicx}	
\usepackage{amsmath}	

\usepackage{amssymb}	
\usepackage{cleveref}   
\usepackage{ulem}
\usepackage{multirow}

\begin{document}

\title{Revisiting the Transit Timing and Atmosphere Characterization of the Neptune-mass Planet HAT-P-26~b}

\author[0000-0001-7234-7167]{Napaporn A-thano}
\affiliation{Department of Physics and Institute of Astronomy, National Tsing-Hua University, Hsinchu 30013, Taiwan}
\email{napaporn@gapp.nthu.edu.tw}

\author[0000-0003-3251-3583]{Supachai Awiphan}
\affiliation{National Astronomical Research Institute of Thailand, 260 Moo 4, Donkaew, Mae Rim, Chiang Mai, 50180, Thailand}
\email{supachai@narit.or.th}

\author[0000-0001-7359-3300]{Ing-Guey Jiang}
\affiliation{Department of Physics and Institute of Astronomy, National Tsing-Hua University, Hsinchu 30013, Taiwan}
\email{jiang@phys.nthu.edu.tw}

\author[0000-0002-1743-4468]{Eamonn Kerins}
\affiliation{Jodrell Bank Centre for Astrophysics, University of Manchester, Oxford Road, Manchester, M13 9PL, UK}

\author[0000-0003-1143-0877]{Akshay Priyadarshi}
\affiliation{Jodrell Bank Centre for Astrophysics, University of Manchester, Oxford Road, Manchester, M13 9PL, UK}

\author[0000-0003-0356-0655]{Iain McDonald}
\affiliation{Jodrell Bank Centre for Astrophysics, University of Manchester, Oxford Road, Manchester, M13 9PL, UK}
\affiliation{Department of Physical Sciences, The Open University, Walton Hall, Milton Keynes, MK7 6AA, UK}

\author[0000-0001-8657-1573]{Yogesh C. Joshi}
\affiliation{Aryabhatta Research Institute of Observational Sciences (ARIES), Manora Peak, Nainital 263001, India}

\author{Thansuda Chulikorn}
\affiliation{Department of Physics, Faculty of Science, Chulalongkorn University, 254 Phayathai Road, Pathumwan, Bangkok 10330, Thailand}

\author{Joshua J. C. Hayes}
\affiliation{Jodrell Bank Centre for Astrophysics, University of Manchester, Oxford Road, Manchester, M13 9PL, UK}

\author{Stephen Charles}
\affiliation{Jodrell Bank Centre for Astrophysics, University of Manchester, Oxford Road, Manchester, M13 9PL, UK}

\author[0000-0003-0728-1493]{Chung-Kai Huang}
\affiliation{Institute of Astronomy and Astrophysics, Academia Sinica, No.1, Sec. 4, Roosevelt Rd, Taipei 10617, Taiwan}

\author{Ronnakrit Rattanamala}
\affiliation{PhD Program in Astronomy, Department of Physics and Materials Science,\\ Faculty of Science, Chiang Mai University, Chiang Mai, 50200, Thailand}
\affiliation{Department of Physics and General Science, Faculty of Science and Technology, \\ Nakhon Ratchasima Rajabhat University, Nakhon Ratchasima, 30000, Thailand}

\author[0000-0001-8677-0521]{Li-Chin Yeh}
\affiliation{Institute of Computational and Modeling Science, National Tsing-Hua University, Hsinchu 30013, Taiwan}

\author{Vik S Dhillon}
\affiliation{Department of Physics and Astronomy, University of Sheffield, Sheffield, S3 7RH, UK}
\affiliation{Instituto de Astrofísica de Canarias, E-38205 La Laguna, Tenerife, Spain}



\begin{abstract}
We present the transit timing variation (TTV) and planetary atmosphere analysis of the Neptune-mass planet HAT-P-26~b. We present a new set of 13 transit light curves from optical ground-based observations and combine them with light curves from the Wide Field Camera 3 (WFC3) on the Hubble Space Telescope (HST), Transiting Exoplanet Survey Satellite (TESS), and previously published ground-based data. We refine the planetary parameters of HAT-P-26 b and undertake a TTV analysis using 33 transits obtained over seven years. The TTV analysis shows an amplitude signal of 1.98 $\pm$ 0.05 minutes, which could result from the presence of an additional $\sim0.02~M_{\textup{Jup}}$ planet at the 1:2 mean-motion resonance orbit. Using a combination of transit depths spanning optical to near-infrared wavelengths, we find that the atmosphere of HAT-P-26~b contains $2.4^{+2.9}_{-1.6}$\% of H$_2$O with a derived temperature of $590^{+60}_{-50}$ K.
\end{abstract}



\section{Introduction} 
\label{sec:intro}

Over the last few decades, the study of exoplanetary systems has grown rapidly, as seen from the number of discovered planets and dedicated surveys. Of the more than 5000 discovered exoplanets so far, about 3000 transiting planets have been discovered\footnote{From NASA exoplanet archive: \texttt{https://exoplanetarchive.ipac.caltech.edu/}.} by several different surveys, such as \textit{Kepler} \citep{borucki2005}, the Transiting Exoplanet Survey Satellite \citep[TESS,][]{ricker2014}, The Wide-Angle Search for Planets \citep[WASP,][]{pol2006,smi2014}, the  Hungarian-made Automated Telescope Network  \citep[HATNet,][]{bak2004,bak2009}, the Kilodegree Extremely Little Telescope \citep[KELT,][]{pep2007} survey, and the Next Generation Transit Survey \citep[NGTS,][]{whe2018}. In addition to the discovery of thousands of exoplanets, the transit technique can also be used to search for additional planets in the system via Transit Timing Variations \citep[TTV;][]{agol2005,agol2018}, and to characterize the compositions of planetary atmospheres via transmission spectroscopy \citep{seager2000,seager2010}.

Current and future detection and atmospheric characterisation missions, including TESS, JWST \citep{pontoppidan2022}, the PLAnetary Transits and Oscillations survey \citep[PLATO,][]{2014ExA....38..249R} and Atmospheric Remote-sensing Infrared Exoplanet Large-survey \citep[ARIEL,][]{2018ExA....46..135T} herald a new era for exoplanetary research. TESS provides continuous, multi-epoch, high-precision light curves, which alone can be used to search for short-term TTVs ($<5$~years). Since 2018, TESS has detected the TTV signals of a number of planets, including two new detections; AU~Mic~c \citep{wittrock2022} and TOI-2202~c \citep{trifonov2021}. 

For exoplanetary atmospheres, the Wide Field Camera 3 (WFC3) on the Hubble Space Telescope (HST) has been used for the detailed study of a number of exoplanets ranging from hot Jupiters to Neptune-sized planets and super-Earths \citep{kreidberg2014,tsiaras2016,burt2021,edwards2021,brande2022,glidic2022}. Since the commencement of science operation in mid-2022, JWST has been used to study the chemical composition of exoplanetary atmospheres in the near-infrared. From the JWST Early Release Observations (ERO) program \citep{pontoppidan2022}, observations from several JWST instruments have revealed the atmospheric compositions of several exoplanets \citep[e.g., WASP-39~b;][]{rustamkulov2022}. 

Whilst HST, Kepler, TESS, JWST, PLATO and ARIEL are all designed to deliver high-quality data from space of exoplanets’ physical and chemical properties, ground-based observations remain critical for long-term monitoring of lightcurve behaviour. The Spectroscopy and Photometry of Exoplanet Atmospheres Research Network (SPEARNET) is a long-term statistical study of the atmospheres of hot transiting exoplanets using transmission spectroscopy. Its observations are supported by a globally distributed heterogeneous network of optical and infrared telescopes with apertures from 0.5 to 3.6 meters, which can be combined with archival data from both ground- and space-based surveys. Our new transit-fitting code, \texttt{TransitFit} \citep{hay2021}, is designed for use with heterogeneous, multi-wavelength, multi-epoch and multi-telescope observations of exoplanet hosts and to fit global parametric models of the entire dataset.

Since 2015, the SPEARNET has monitored transits of HAT-P-26~b, which is a Neptune-mass planet orbiting a host K1 dwarf HAT-P-26 ($V$~=~11.74) with a period of 4.234~days \citep{hartman2011}. The stellar and planetary parameters of the HAT-P-26 system are given in \Cref{tab:HATP26-Properties}. Transmission spectra of HAT-P-26~b were first studied by \citet{steven2017}. Using the observations from \textit{Magellan} and \textit{Spitzer}, they reported that HAT-P-26~b is likely to have high metallicity, with a cloud-free upper atmosphere containing water and a 1000 Pa cloud deck. \citet{wak2017} obtained observations from HST and Spitzer Space Telescopes, which showed a high-significance detection of H$_2$O and a metallicity approximately 4.8 times solar abundance. 

\citet{macdonald2019} combined previous HST and Spitzer data for HAT-P-26~b with ground-based spectroscopic observations from the \textit{Magellan} Low Dispersion Survey Spectrograph 3 \citep[LDSS-3C,][]{steven2017}. From the study, H$_2$O was detected with an abundance of $1.5\%$  and O/H with an abundance 18.1 times solar. They also reported evidence for metal hydrides in the spectra with $>4\sigma$ confidence with the potential candidates identified as TiH, CrH, or ScH. The presence of metal hydrides in the atmosphere requires extreme conditions, such as the vertical transportation of material from the deep atmosphere or solid planetesimals containing heavy elements impacting the planet and dissolving the elements into the He/H$_2$ envelope through shocks and fireballs. 

Besides the study of transmission spectroscopy, HAT-P-26~b was examined for TTVs by \citet{von2019}. They performed follow-up photometric observation with the 2.15~m Jorge Sahade Telescope, Argentina, as well as a 1.2~m robotic telescope (STELLA) and the 2.5~m Nordic Optical Telescope, both located in the Canary Islands. The observed transits showed a $\sim270$-epochs periodic timing variation with an amplitude of $\sim4$ minutes, which might be caused by the third body in the system. 

In this work, we present new ground-based SPEARNET multi-band photometric follow-up observations of 13 transits of HAT-P-26~b. These data are combined with TESS,  HST, and available published photometric data to constrain the planetary physical parameters, investigate the planetary TTV signal, and constrain the atmospheric model. Our observational data are presented in Section~\ref{sec:observation}. The light-curve analysis is described in Section~\ref{sec:LCModeling}. A new linear ephemeris and a frequency study of TTVs is presented in Section~\ref{sec:ttv}. In Section~\ref{sec:atmosphere}, the atmospheric composition of HAT-P-26~b is analysed. Finally, the discussion and conclusion are in Section~\ref{sec:conclusion}.

\begin{table*}
\begin{center}
\caption{Summary of HAT-P-26 system's properties from \citet{hartman2011}.}
\label{tab:HATP26-Properties}          
\small\addtolength{\tabcolsep}{-2pt}
\begin{tabular}{lc}
\toprule
Parameter  &  Value  \\ 
\hline
\multicolumn{2}{c}{Stellar Parameters}\\
\hline
$M_\star$  & 0.82 $\pm$ 0.03~M$_\odot$  \\
$R_\star$  & 0.79~$\pm$~0.01~R$_\odot$  \\
$T_*$ & 5079~$\pm$~88~K  \\
log$_{g\star}$     & 4.56~$\pm$~0.06~cgs  \\
Metallicity [Z$_{*}$] & $-0.04~\pm~0.08$  \\
\hline
\multicolumn{2}{c}{Planetary Parameters}   \\
\hline
$M_p$ & 0.059~$\pm$~0.007 M$_{\textup{Jup}}$  \\
$R_p$ & $0.565^{+0.072}_{-0.032}$ R$_{\textup{Jup}}$   \\
T$_\textup{eq}$ & $1001^{+66}_{-37}$ K   \\
$\rho_{p}$ & 0.40~$\pm$~0.10~g.cm$^{-3}$ \\
$P$ (days)  &  4.234516 $\pm$ 2 $\times$ 10$^{-5}$ \\
$i$ (deg)        &   $88.6^{+0.5}_{-0.9}$  \\
{$a/R_\ast$}     &   13.06  $\pm$	0.83  \\
\hline
\end{tabular}
\end{center}
\end{table*}

\section{Observational Data}
\label{sec:observation}
Since the discovery of HAT-P-26~b in 2011, the planetary system has been monitored by a number of campaigns, as discussed in Section~\ref{sec:intro}. In this work, we present the data from our observations (13 transit light curves) and previously published data (69 transit light curves). The details of each observational data set are described below.

\subsection{SPEARNET Observations and Data Reduction}
\label{subsec:SPEARNETdata}

\begin{figure}
    \centering
    \includegraphics[width=1\textwidth,page=1]{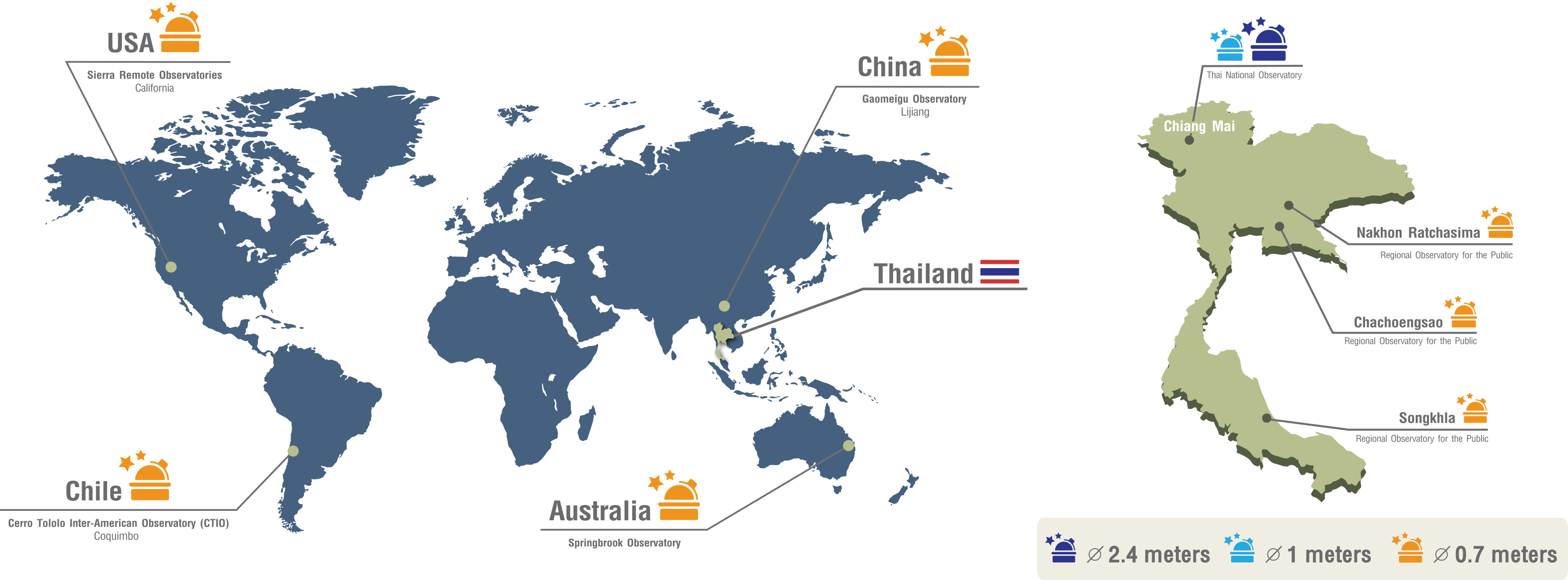}
    \caption{The locations and sizes of the SPEARNET telescope network.} 
    \label{fig:TRT}
\end{figure}

Between March 2015 and May 2022, photometric follow-up observations of HAT-P-26~b were obtained using the SPEARNET telescopes network (\Cref{fig:TRT}). Time-series photometry of thirteen transits, including eight full and five partial transits, were obtained. The observation log is given in \Cref{tab:log}. The facilities used to obtain our data were as follows: 

\begin{table*}
\begin{center}
\caption{Observation details of HAT-P-26~b's transits within using the telescopes within the SPEARNET. Epoch=0 is the transit on 2010 April 18.}
\label{tab:log}          
\small\addtolength{\tabcolsep}{-2pt}
\begin{tabular}{lcccccccc}
\toprule
\multirow{2}{*}{Observation Date}  & \multirow{2}{*}{Epoch} & \multirow{2}{*}{Telescope}  & \multirow{2}{*}{Filter} & \multirow{2}{*}{Exposure time (s)} & Number & Total Duration of & \multirow{2}{*}{PNR (\%)} & Transit \\
  &   &   &   &  & of Images & Observation (hr) &   &  coverage \\
\hline
2015 March 05   & 421   & 2.4-m TNT  & $i'$ & 1.90   & 3892 & 2.49   & 0.09   & Egress only  \\
2015 March 22   & 425   & 2.4-m TNT  & $i'$ & 2.47    & 6683 & 4.92  & 0.12   & Full   \\
2016 February 11  & 502  & 2.4-m TNT & $g'$ & 9.23    & 1574 & 4.19  & 0.07   & Full  \\  
2017 March 15   & 596   & 0.7-m TRT-GAO   & $R$  & 40 & 235  & 3.65 & 0.19  & Ingress only  \\  
2017 March 15   & 596   & 0.5-m TRT-TNO   & $I$  & 30 & 270  & 3.13 & 0.17  & Ingress only   \\ 
2018 March 27   & 685  & 2.4-m TNT  & $g'$  & 4.53    & 4465 & 5.77  & 0.35    & Full   \\ 
2018 March 27   & 685  & 0.5-m TRT-TNO  & $R$  & 40   & 216 & 2.93  & 0.21    & Full    \\ 
2018 April 13   & 689  & 2.4-m TNT  & $z'$  & 2.68 & 4590 & 4.51   & 0.17   & Full   \\  
2019 March 05   & 766  & 2.4-m TNT  & $r'$  & 2.98 & 5493 & 4.71  & 0.10   & Full  \\  
2019 April 25  & 778  & 2.4-m TNT  & $z'$   & 4.86 & 2481  & 4.12  & 0.14  & Egress only  \\ 
2022 March 23  & 1029  & 0.7-m TRT-SRO  & $R$    & 30  & 353 & 4.01   & 0.24   & Full    \\ 
2022 May 13    & 1041  & 0.7-m TRT-SRO  & $I$    & 30  & 273 & 3.32  & 0.40   & Full     \\  
2022 May 30    & 1045  & 0.7-m TRT-SRO  & $R$    & 30  & 226 & 2.10 & 0.25   & Egress only  \\  
\hline
\end{tabular}
\end{center}
{\textbf{Note}: PNR is the photometric noise rate \citep{Fulton2011}.}
\end{table*}

\begin{enumerate}
    \item \emph{2.4-m Thai National Telescope (TNT)} located at the Thai National Observatory (TNO), Thailand. During 2015-2019, five full transits and two partial transits of HAT-P-26~b were obtained by the TNT. The observations were conducted using ULTRASPEC \citep{dhillon2014}, a high-speed frame-transfer EMCCD 1024 $\times$ 1024 pixels camera, with a field-of-view of 7.68 $\times$ 7.68 arcmin$^{2}$.  
    
    \item \emph{0.5-m Thai Robotic Telescope located at TNO (TRT-TNO), Thailand.} We observed one full transit and one partial transit of HAT-P-26~b between 2017 and 2018 with the Schmidt-Cassegrain TRT-TNO. (currently, the facility is upgraded to a 1-m telescope). The observations were performed using an Apogee Altra U9000 3056 $\times$ 3056 pixels CCD camera. The field of view is about 58 $\times$ 58 arcmin$^{2}$.
     
    \item \emph{0.7-m Thai Robotic Telescope at the Gao Mei Gu Observatory (TRT-GAO), China.} One partial transit of HAT-P-26~b was obtained by the TRT-GAO in 2017. TRT-GAO is equipped with an Andor iLon-L 936, with a 2048 $\times$ 2048 pixels CCD camera. The field of view is 20.9 $\times$ 20.9 arcmin$^{2}$. 
    
    \item \emph{0.7-m Thai Robotic Telescope at the Sierra Remote Observatories (TRT-SRO), USA.} In 2022, the TRT-SRO obtained two full and one partial transit. We observed HAT-P-26~b with the Andor iKon-M 934 1024 $\times$ 1024 pixels CCD camera. The field of view is 10 $\times$ 10 arcmin$^{2}$.
\end{enumerate} 

All the science images of HAT-P-26~b were pre-processed using standard tasks from {\tt\string IRAF}\footnote{IRAF is distributed by the National Optical Astronomy Observatories, which are operated by the Association of Universities for Research in Astronomy, Inc., under a cooperative agreement with the National Science Foundation. For more details, \texttt{http://iraf.noao.edu/}} \citep{tody1986,tody1993}. Astrometric calibrations were performed using {\tt\string Astrometry.net} \citep{lang2010}, and aperture photometry was performed by {\tt\string source extractor} \citep{berlin1996}. We use {\tt\string mag\_auto}, which is Kron-like automated scaled aperture magnitude, with a Kron factor of 2.5 and a minimum radius of 3.5. Reference stars were selected from nearby stars that were within $\pm$ 3 magnitudes of HAT-P-26 and that did not exhibit strong brightness variation. The sigma clipping algorithm, with a 5-sigma threshold, was employed to remove the outlier points in the light curves. To produce the light curves, the flux of HAT-P-26 was divided by the sum of the flux from the selected reference stars. Image time stamps were converted to Barycentric Julian Date in Barycentric Dynamical Time (BJD$_\textup{TDB}$) using {\tt\string barycorrpy} \citep{Kano2018}. The normalized light curves are available in a machine-readable form in \Cref{tab:lightcurve}.

\subsection{Existing Ground-based Data}

We used 16 additional light curves from two previous ground-based studies. Firstly, five $i'$-band transits of HAT-P-26~b were obtained using the KeplerCam on the FLWO 1.2 m telescope obtained by \citet{hartman2011}\footnote{Download from the CDS: \texttt{https://cdsarc.cds.unistra.fr/viz-bin/cat/J/ApJ/728/138}}. Secondly, we used 11 Cousins-$R$ transits obtained by \citet{von2019} using the 2.15 m Jorge Sahade Telescope at the Complego Astron\'omico El Leoncito (CASLEO), the 2.5 m Nordic Optical Telescope (NOT) at La Palma, Spain, and the 1.2 m STELLA at Tenerife, Spain\footnote{Download from the CDS: \texttt{https://cdsarc.cds.unistra.fr/viz-bin/cat/J/A+A/628/A116}}. These are summarized in \Cref{tab:log_achived}. Combining these data with our observations of 13 transits, we use ground-based photometry from 29 transits obtained over a span of 20 years within six broad photometric bands. 

\begin{table*}
\begin{center}
\caption{Summary of HAT-P-26~b's transits light curves taken from Ground-based Achieved Data.}
\label{tab:log_achived}          
\small\addtolength{\tabcolsep}{-2pt}
\begin{tabular}{lcc}
\toprule
Observation Date  &  Telescope  & Filter  \\
\hline
\multicolumn{3}{c}{\citet{hartman2011}}\\
\hline
2010 January 05$^*$ & KeplerCam/the FLWO 1.2-m & $i'$ \\
2010 March 31$^*$ & KeplerCam/the FLWO 1.2-m & $i'$  \\
2010 April 04 & KeplerCam/the FLWO 1.2-m & $i'$  \\
2010 May 08 & KeplerCam/the FLWO 1.2-m & $i'$  \\
2010 May 25 & KeplerCam/the FLWO 1.2-m & $i'$  \\
\hline
\multicolumn{3}{c}{\citet{von2019}}   \\
\hline
2015 March 30 & the 2.15-m CASLEO  & $R$  \\
2015 April 12 & the 2.5-m NOT & $R$   \\
2015 April 16 & the 2.15-m CASLEO & $R$   \\
2015 May 20$^*$ & the 2.5-m NOT & $R$   \\
2015 June 06 & the 2.5-m NOT & $R$   \\
2015 June 23 & the 2.5-m NOT & $R$   \\
2016 May 14$^*$ & the 2.15-m CASLEO & $R$   \\
2017 May 13$^*$ & the 2.15-m CASLEO & $R$   \\
2017 May 30 & the 2.15-m CASLEO & $R$   \\
2017 June 16$^*$ & the 2.15-m CASLEO & $R$  \\
2018 July 01$^*$ & the 1.2-m STELLA & $R$  \\
\hline
\end{tabular}
\end{center}
{\textbf{Note}: $^*$ Only part of the transit was observed.}
\end{table*}

\subsection{HST WFC3 Grism Data}
In addition to ground-based observations, HST observed three transits of HAT-P-26~b using WFC3 \citep{wak2017}. Two transits were observed using the G141 grism (1.1 to 1.7 $\mu$m) on 2016 March 12 and 2016 May 02. Another transit was observed using the G102 grism (0.8 to 1.1 $\mu$m) on 2016 October 16. 

The raw spectra were reduced using the \texttt{Iraclis} package, a Python package for the WFC3 spectroscopic reduction pipeline \citep{tsiaras-waldmann2016,tsiaras2016}\footnote{Downloaded from Exo.MAST: \texttt{https://exo.mast.stsci.edu/}}. The HST data from the G141 grism spectra were binned into 18 wavelength bins, while the G102 grism spectra were binned into 14 wavelength bins. In total, 50 light curves were obtained from HST/WFC3. We discarded the data from the first orbit of each visit and the first exposure of each orbit as the data exhibit a stronger wavelength-dependent ramp during these epochs. 
 
\subsection{TESS Data}
TESS observed three transit light curves of HAT-P-26 b in Sector 50 (2022 March-April). We used the Pre-Search Data Conditioning (PDC) light curves \citep{smith2017,smith_pdc2017}, which were the calibrated light curves from the Science Processing Operation Center (SPOC) pipeline \citep{jenkins2016}\footnote{Downloaded from the Mikulski Archive for Space Telescopes: \texttt{https://archive.stsci.edu/}}. The dilution and background-corrected PDCSAP light curves from the SPOC pipeline are used in this work.

\begin{table*}
\begin{center}
\caption{A sample of the detrended and normalised photometry for HAT-P-26~b using the telescopes within SPEARNET. The transits were all detrended with a second-order polynomial function in \texttt{TransitFit}. Epoch~=~0 is the transit on 2010 April 18.} 
\label{tab:lightcurve}          
\small\addtolength{\tabcolsep}{-2pt}
\begin{tabular}{lcccc}
\toprule
\multirow{2}{*}{Epoch} & \multirow{2}{*}{BJD} & \multirow{2}{*}{Normalized Flux} & Normalized flux   \\
      &     &                 & uncertainty       \\
\hline
421	& 2457087.35923	& 0.995	& 0.004	\\
	& 2457087.35925	& 0.996	& 0.005	\\
	& 2457087.35927	& 0.998	& 0.005	\\
	& 2457087.35934	& 1.000	& 0.004	\\
	& 2457087.35936	& 0.996	& 0.004	\\
	& ...	& ...	& ...	\\
\hline							
425	& 2457104.23686	& 0.990	& 0.004	\\
	& 2457104.23689	& 1.002	& 0.004	\\
	& 2457104.23694	& 0.990	& 0.004	\\
	& 2457104.23700	& 1.011	& 0.004	\\
	& 2457104.23703	& 1.002	& 0.004	\\
	& ...	& ...	& ...	\\
\hline							
502	& 2457430.29378	& 1.002	& 0.002	\\
	& 2457430.29388	& 1.002	& 0.002	\\
	& 2457430.29420	& 0.998	& 0.002	\\
	& 2457430.29431	& 1.002	& 0.002	\\
	& 2457430.29452	& 0.999	& 0.002	\\
	& ...	& ...	 & ...	\\
\hline
... & ...   & ...    & ...        \\
\hline
\end{tabular}
\end{center}
{\textbf{Note}: The full table is available in machine-readable form.}
\end{table*}

\section{Light-Curve Modeling}
\label{sec:LCModeling}
HAT-P-26~b has been observed by several observing campaigns, which report subtly different planetary physical parameters. The differences can arise from different modeling assumptions, such as the treatment of limb darkening. In this present study, the physical parameters of HAT-P-26~b are reanalyzed using the \texttt{TransitFit} (Version 3.0.9), a Python package that can simultaneously fit multi-filter, multi-epoch exoplanet transit observations \citep{hay2021}. \texttt{TransitFit} models transits using \texttt{batman} \citep{kreidberg2015} and  performs fitting using the dynamic nested-sampling routine from \texttt{dynesty} \citep{speagle2020}.

The combined ground and space datasets comprise 85 separate light curves spanning a range of epochs and wavelengths. We fit and detrend all of them simultaneously using \texttt{TransitFit}. \texttt{TransitFit} performed nested-sampling retrieval with 1000 live points and a slice sampling of 10. During the retrieval, each transit light curve was individually detrended using different detrending functions: for each ground-based observation, we used individual second-order polynomial detrending functions. For the HST/WFC3 data sets, the data were detrended using a model similar to \citep{kreidberg2018}, specifically
\begin{equation}
    F_{sys} = (S + v_{1}t_{visit} + v_{2}t^{2}_{visit})(1 - e^{-at_{orb}-b}) \ ,
\end{equation}
where $F_\textup{sys}$ is the signal from the systematics, and $S$ = 1 and $s$ for forward and reverse scans, respectively. The parameters $s$, $v_{1}$, $v_{2}$, $a$, and $b$ are all detrending coefficients, where $s$, $a$, and $b$ account for the ramp-up systematic across all the light curves, whilst $v_{1}$ and $v_{2}$ are the second-order polynomial detrending functions used to model the visit-long trends. The astrophysical signal ($F_\textup{sig}$) can be obtained by the division of the observed flux ($F_\textup{obs}$) and the systematic signal ($F_\textup{sys}$). The HST detrending function was defined as a custom detrending function in \texttt{TransitFit}. The normalized light curves with their observational uncertainties are available in a machine-readable form in \Cref{tab:lightcurve}. 

\begin{figure*}[htb]
\centering
    \includegraphics[width=0.85\textwidth]{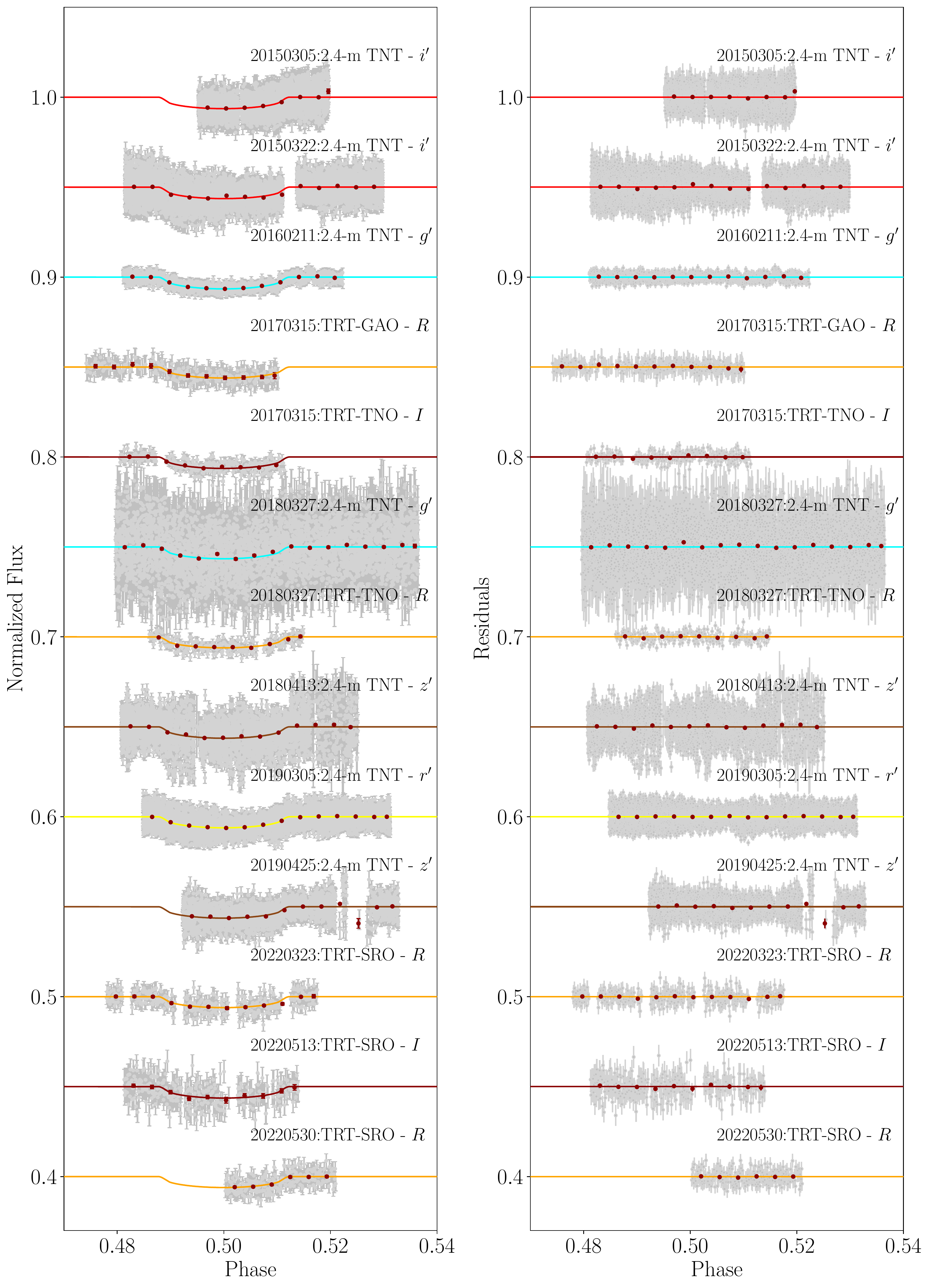} 
    \caption{{\it Left panels:} Normalized, phased-folded HAT-P-26~b transit light curves observed using the SPEARNET telescope network (gray dots) with the best-fitting model from \texttt{TransitFit} (solid lines). The red dots show the light curves binned into 5 minute intervals. {\it Right panels:} The corresponding residual lightcurves after models are subtracted. Both the light curves and the residuals have arbitrary vertical offsets for clarity.}
    \label{fig:LCs_TNT}
\end{figure*}

\begin{figure*}[htb]
\centering
  \begin{tabular}{cc}
    \includegraphics[width=0.49\textwidth,page=1]{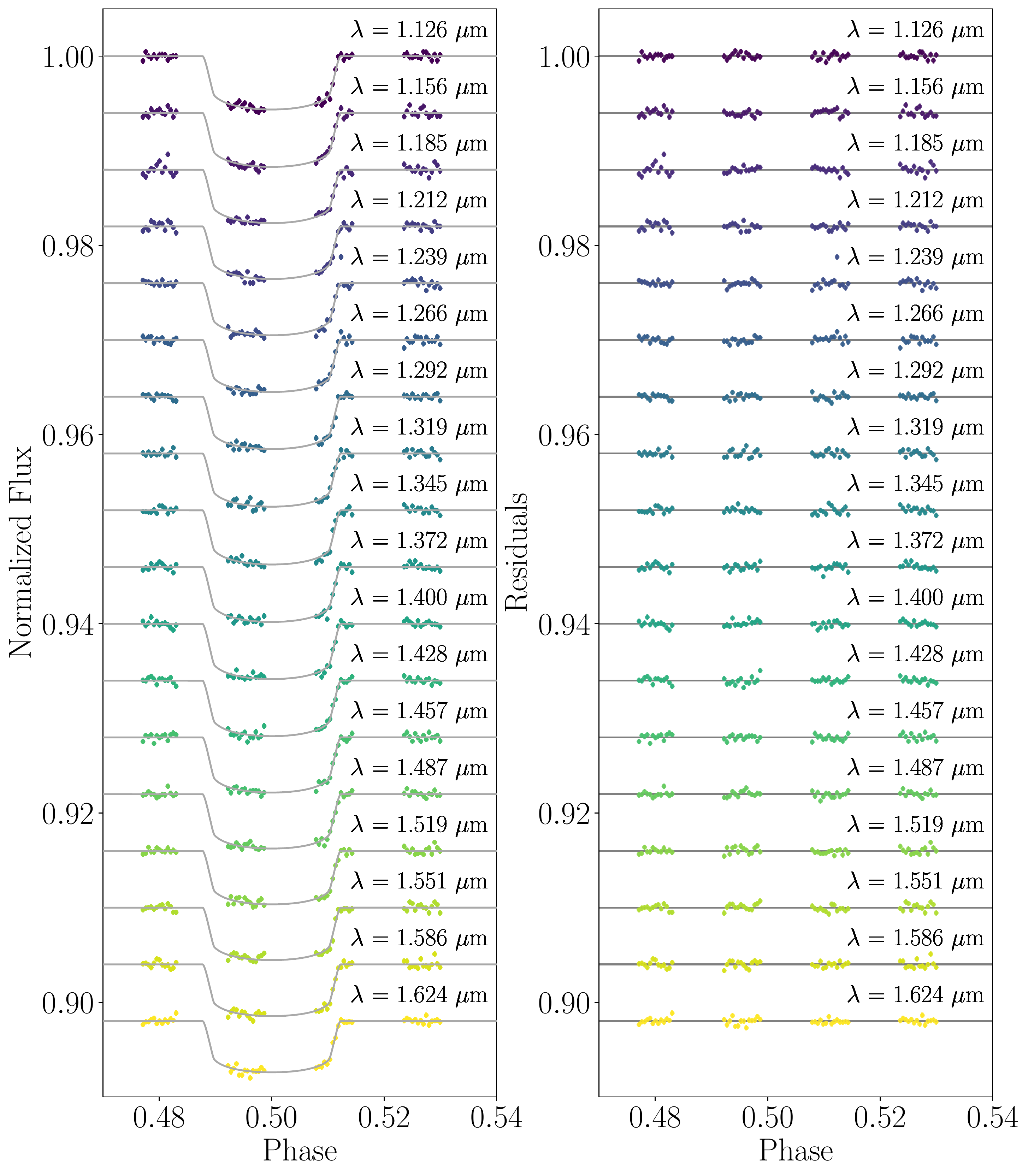} 
    \includegraphics[width=0.49\textwidth,page=1]{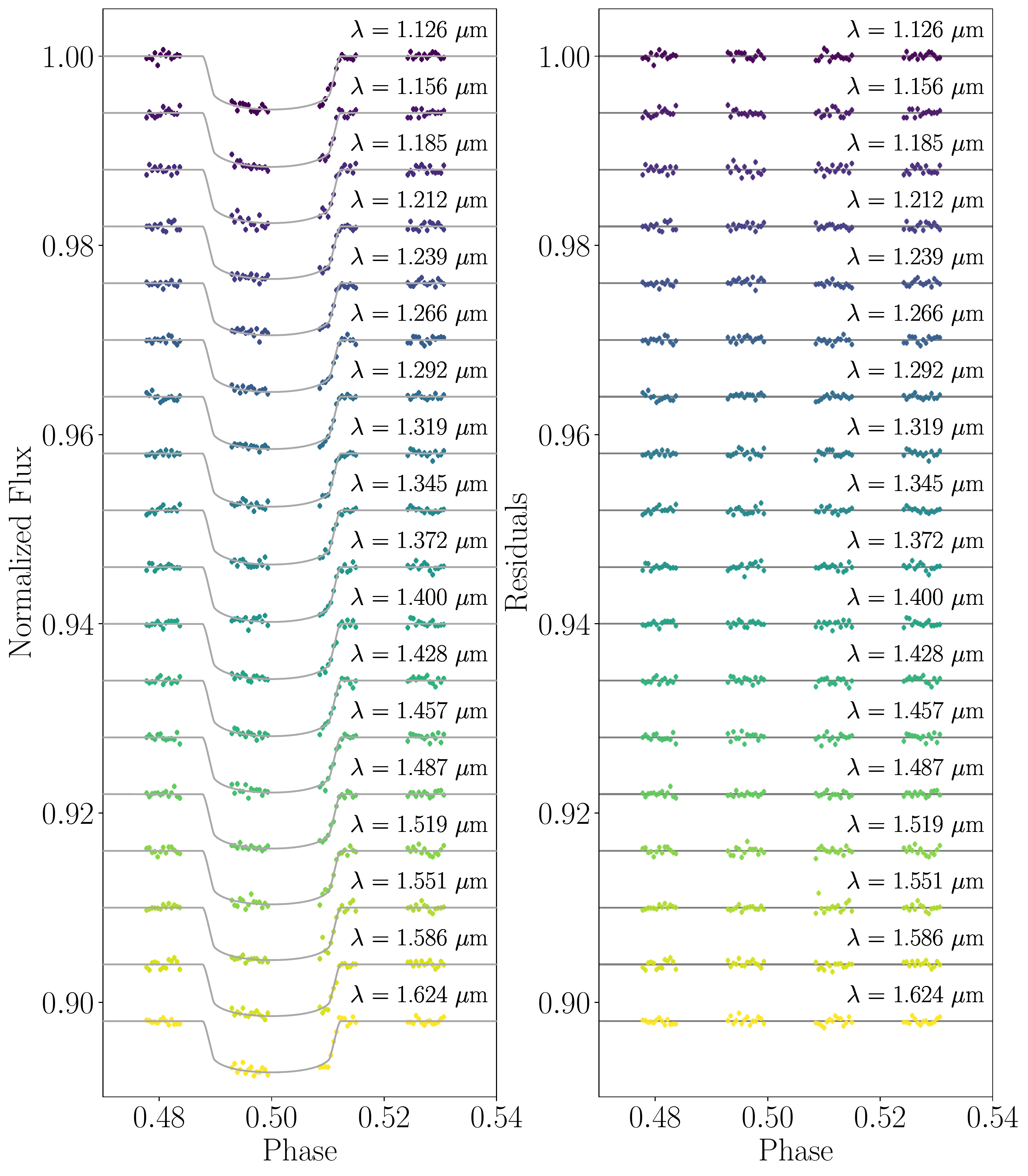} \\
    \includegraphics[width=0.49\textwidth,page=1]{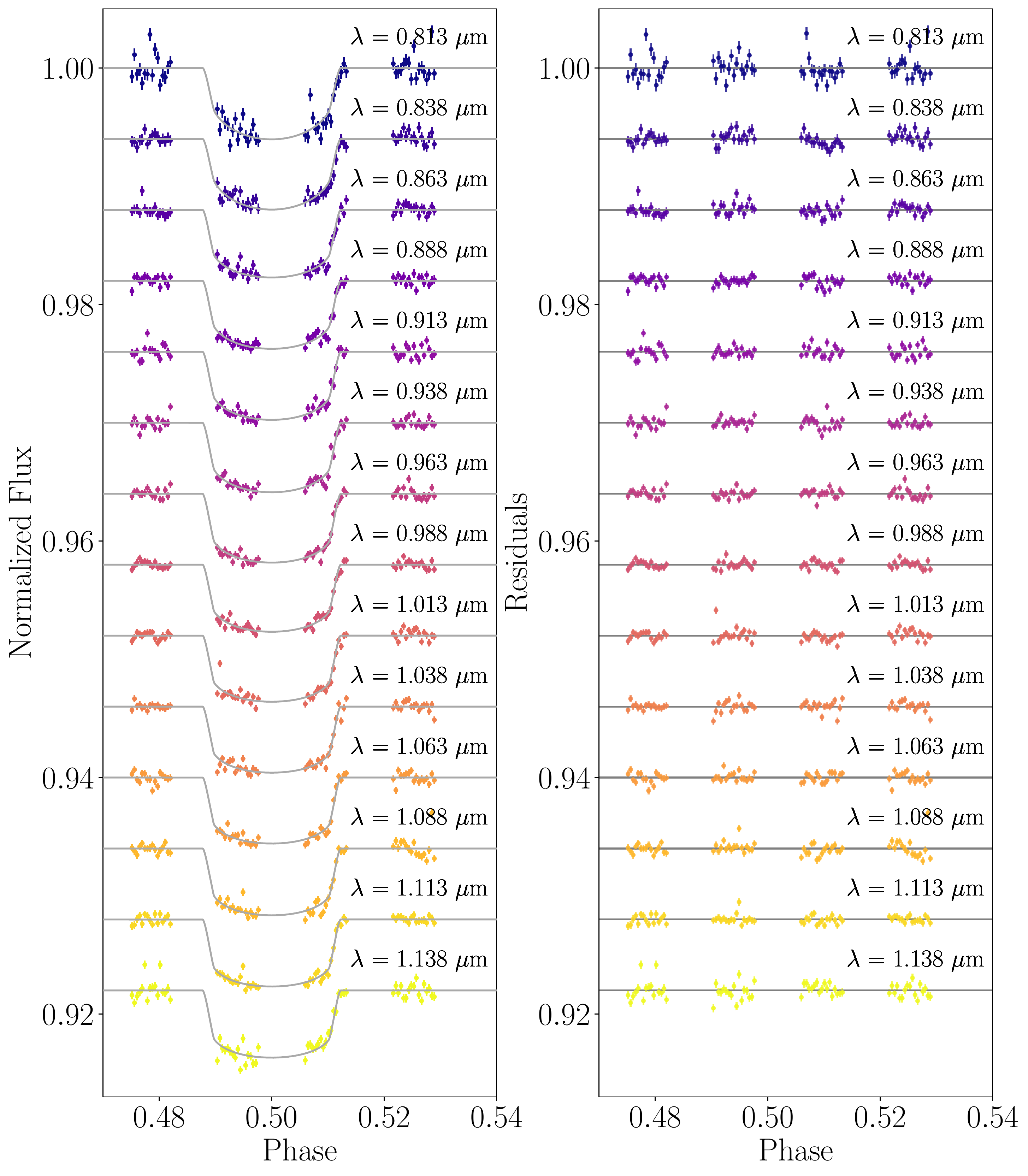} 
    \end{tabular}
    \caption{Normalized, phased-folded HAT-P-26~b transit light curves from HST/WFC3, reduced using the \texttt{Iraclis} package (dots). Three sets of observations are shown, with the light curves and \texttt{TransitFit} models in the left-hand panels, and the corresponding residual differences in the right-hand panels. The top-left, top-right and bottom pairs of panels respectively show the G141 grism observations from 2016 March 12, the G141 grism observations from 2016 May 02, and the G102 prism observations from 2016 October 16 (bottom). The light curves and the residuals have arbitrary offsets for clarity.}
    \label{fig:LCs_HST}
\end{figure*}

\begin{figure*}[htb]
\centering
    \includegraphics[width=0.49\textwidth,page=1]{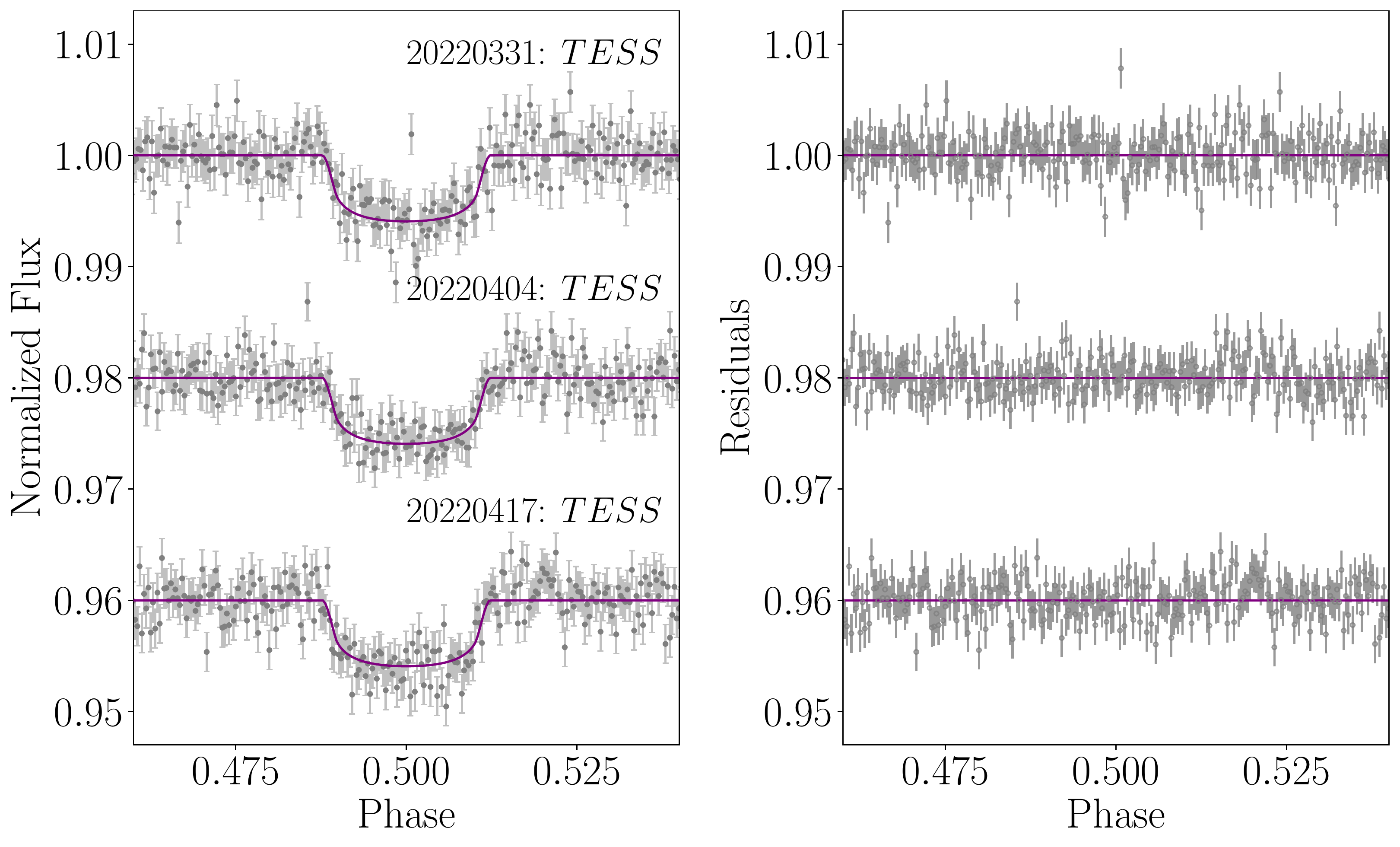} 
    \caption{Normalized, phased-folded HAT-P-26~b transit light curves from the TESS (gray dots, left panels) with the best-fitting model from \texttt{TransitFit} (solid lines). Their corresponding residuals are shown in the right panels. The light curves and the residuals have arbitrary offsets for clarity.}
    \label{fig:LCs_TESS}
\end{figure*}

HAT-P-26 b is assumed to be in a circular orbit. We find a stellar effective temperature for HAT-P-26 of $T_*=4700\pm100$, determined from the Python Stellar Spectral Energy Distribution package\footnote{\url{https://explore-platform.eu/}}, a toolset designed to allow the user to create, manipulate and fit the spectral energy distributions of stars based on publicly available data \citep{mcdonald2009,mcdonald2012,mcdonald2017}. To create the stellar SED, we obtained the available photometry, which consists of the G, G$_\textup{BP}$, and G$_\textup{RP}$ magnitudes from Gaia, the $g', r', i', z', y'$ magnitudes from the Panoramic Survey Telescope and Rapid Response System (Pan-STARRS), the near-ultraviolet (NUV) magnitude from the Galaxy Evolution Explorer (GALEX), the $B, V, g', r', i'$
from the AAVSO Photometric All-Sky Survey (APASS), the $JHK$ magnitudes from the Two Micron All Sky Survey (2MASS), and the W1–W3 magnitudes from the Wide-ﬁeld Infrared Survey Explorer (WISE), also the BT-Settl model atmospheres \citep{allard2011,allard2014} were used.

The host metallicity, $Z_{*} = -0.06\pm0.10$, and surface gravity, $\textup{log} (g_*) = 4.5\pm0.1$, are obtained from the Gaia EDR3 catalogue\footnote{Gaia archive: \texttt{https://archives.esac.esa.int/gaia}}. To fit ground-based and HST light curves, we fixed the orbital period ($P$) of 4.234516 days, which was adopted from \citet{hartman2011}, and used the ability of \texttt{TransitFit} to account for TTVs by using the \texttt{allow\_TTV} function, in order to find the mid-transit time, $T_m$, for each epoch. The parameters of inclination, $i$, semi-major axis $a$, and planet-to-host radius ratio, $R_p/R_{\star}$ were allowed to vary freely. The priors of each fitting parameter: the epoch of mid-transit, $T_0$, together with $i$, $a$ and $R_p$ for each waveband, are given in \Cref{tab:initialpara}. 

The light curves of HAT-P-26 b were phase-folded to center $T_0$ at a phase of 0.5 with their best-fit models, and residuals are shown in \Cref{fig:LCs_TNT,fig:LCs_HST}. The derived planetary parameters for HAT-P-26 b from \texttt{TransitFit} are compared with the results from previous studies in \Cref{tab:outpara}. HAT-P-26 b has $i = 87.82 \pm 0.05$ deg with a host separation of 12.51 $\pm$ 0.07 $R_\ast$ as shown in \Cref{tab:outpara}. These values exhibit a difference of $\sim1\sigma$ compared to the previous published measurements.

The discrepancies between our fitting result and previous measurements in inclination and host separation might be caused by the missing ingress or egress of the transit or the interruption during the transit due to the weather in our ground-based light curves. Therefore, we perform another fitting analysis using the ground-based light curves which have the data during both ingress and egress plus at least 20 minutes out-of-transit baseline. We defined these light curves as ``Full light curves”. The fitting show that the full light curves provide the same planetary parameters within $2\sigma$ of all light curves fitting (\Cref{tab:outpara}). This test ensures that the fitting results are not biased by the inclusion of partial transit light curves. Since there are no significant differences observed between the fitting results obtained from the analysis of all light curves and the full light curves, we focus on the results derived from the all light curve fitting in this study.

\begin{figure*}[htb]
\centering
  \begin{tabular}{cc}
    \includegraphics[width=0.975\textwidth]{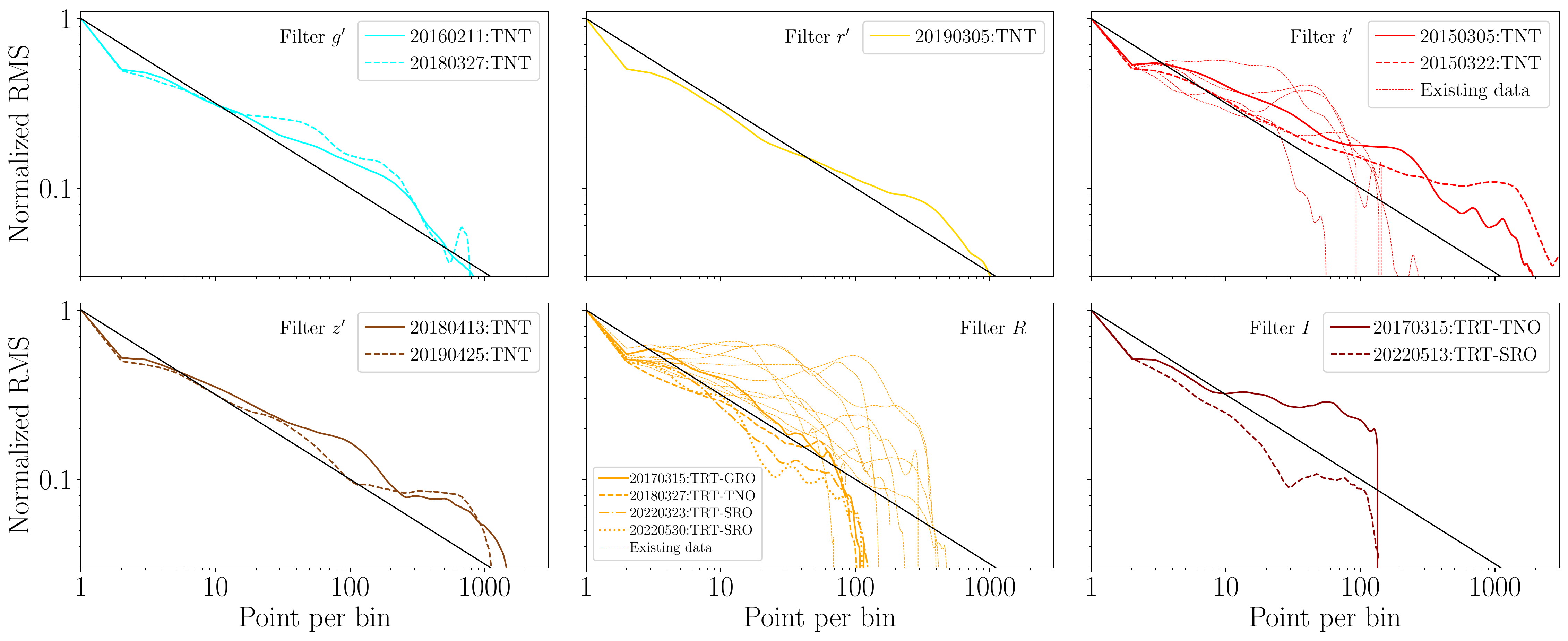}
    \end{tabular}
    \caption{The normalised rms plots for all ground-based residauls light curves from both our observations and existing data fit as a function of the number of points per bin. The observation data are shown in $g'$ (Blue), $r'$ (Yellow), $R$ (Orange), $i'$ (Red), $I$ (Dark-red), $z'$ (Brown).}
    \label{fig:Allan-TNT}
\end{figure*}

\begin{figure*}[htb]
\centering
  \begin{tabular}{cc}
    \includegraphics[width=0.9\textwidth]{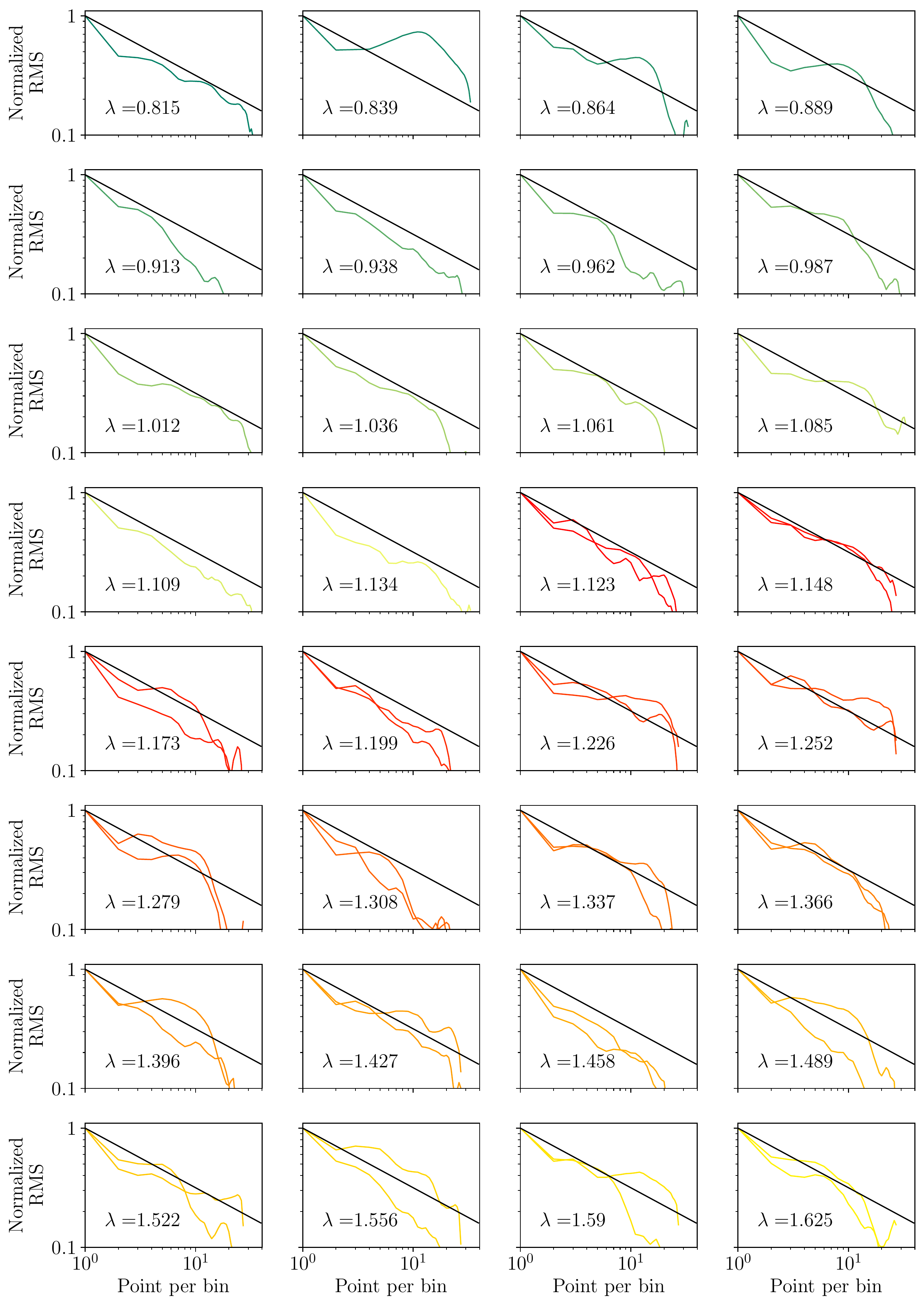} 
    \end{tabular}
    \caption{The normalised rms plots for HST residual light curves fit as a function of the number of points per bin. The observations from HST/WFC3 G102 and HST/WFC3 G141 are shown in green-gradient and red-gradient, respectively.}
    \label{fig:Allan-HST}
\end{figure*}

Furthermore, an exploration into the observed noise within the light curves was undertaken by examining the normalised root-mean-square (rms) behavior when the light curve is binned in time, following the methodology outlined in \citet{kreidberg2018b}. The hypothetical white noise curve should decrease by a factor of $\sqrt{N}$, where $N$ represents the number of points per bin. The normalised rms values were calculated for both the ground-based and HST light curves, and the results are shown in \Cref{fig:Allan-TNT,fig:Allan-HST}. The overall trend observed in the 2.4-m TNT data indicates the presence of white noise. Nevertheless, as the bin size increases, there is evidence of red noise, particularly within the $i'$ filter. For the TRTs data, most of the light curves exhibit characteristics of white noise, except for the $I$ filter data from TRT-TNO, which displays an increasing degree of red noise with larger bin sizes. Regarding the HST/WFC3 data collected from the G102 and G141 grisms, the analysis reveals a prevalence of white noise throughout the dataset. However, in the G102 data, some instances of red noise become apparent in the large bin size of the light curves when binned within the initial wavelength range (0.8-0.9 $\mu$m). Additionally, the publicly available light curves in the $R$-band also show the existence of time-correlated noise. This presence of red noise within these specific data and wavelength bands might potentially be attributed to the quality of data obtained during those observations.

Mid-transit times ($T_m$) for each transit and corresponding epoch number, $E$, are given in \Cref{tab:midtransit} and discussed in Section~\ref{sec:ttv}. The values of $R_p/R_{\star}$ are shown in \Cref{tab:radius-transitDepth-limbdark}. We can now compare the $R_p/R_{\star}$ values obtained from \texttt{TransitFit} with those from previous studies. The transit depths obtained from the \texttt{TransitFit} exhibit variations at the $\sim$5-sigma level, which can be explained by wavelength-dependent variations of the atmospheric transmission spectrum. For instance, in the $i'$ filter, our observation of $R_p/R_{\star} = 0.0716 \pm 0.0001$ is consistent within a 2-sigma range of the value reported by \citet{hartman2011} (0.0737 $\pm$ 0.0012). However, in the $R$ filter, we found a shallower transit depth ($R_p/R_{\star} = 0.0698 \pm 0.0002$) compared to the measurement provided by \citet{von2019} (0.07010 $\pm$ 0.00016). For the HST filters, the $R_p/R_{\star}$ values from \texttt{TransitFit} are consistent with the values provided by \citet{wak2017}. In the case of TESS, the fitted value for $R_p/R_{*}$ is calculated to be $0.0711 \pm 0.0007$ which aligns with results obtained at other wavelengths. These transit depths are used for the atmospheric modelling in Section~\ref{sec:atmosphere}.

For limb-darkening coefficients, the quadratic limb-darkening coefficients from \citet{hartman2011} were $u_0$ = 0.386 and $u_1$ = 0.258, adopted from the tabulations by \citet{claret2004} for the $i'$-filter, based on a stellar temperature of $T_{*} = 5079 \pm 88$ K and metallicity of $Z_{*} = -0.04\pm0.08$. Similar quadratic limb-darkening coefficients in \citet{von2019} were taken from the $R$-filter tabulated values of \citet{claret2000} as $u_0$ = 0.514 and $u_1$ = 0.218, based on $T_{*} = 5000$ K and $Z_{*} = 0$. Due to the broad range of wavebands analysed in this work, the coupled fitting mode in \texttt{TransitFit} was used to determine the limb-darkening coefficients (LDCs) for each filter. The LDC fitting is conditioned using priors generated by the Limb Darkening Toolkit \citep[LDTk,][]{parviainen2015} for each filter response, based on the PHOENIX\footnote{PHOENIX: \texttt{http://phoenix.astro.physik.uni-goettingen.de/}} stellar atmosphere models \citep{husser2013}. Our previously determined host star parameters, including the $T_*$, $\log(g)$ and $Z_{*}$, are adopted for the LDC calculations. The LDCs for different filters from the coupled fitting mode are given in \Cref{tab:radius-transitDepth-limbdark}. 

\begin{figure*}[htb]
\centering
    \includegraphics[width=0.42\textwidth]{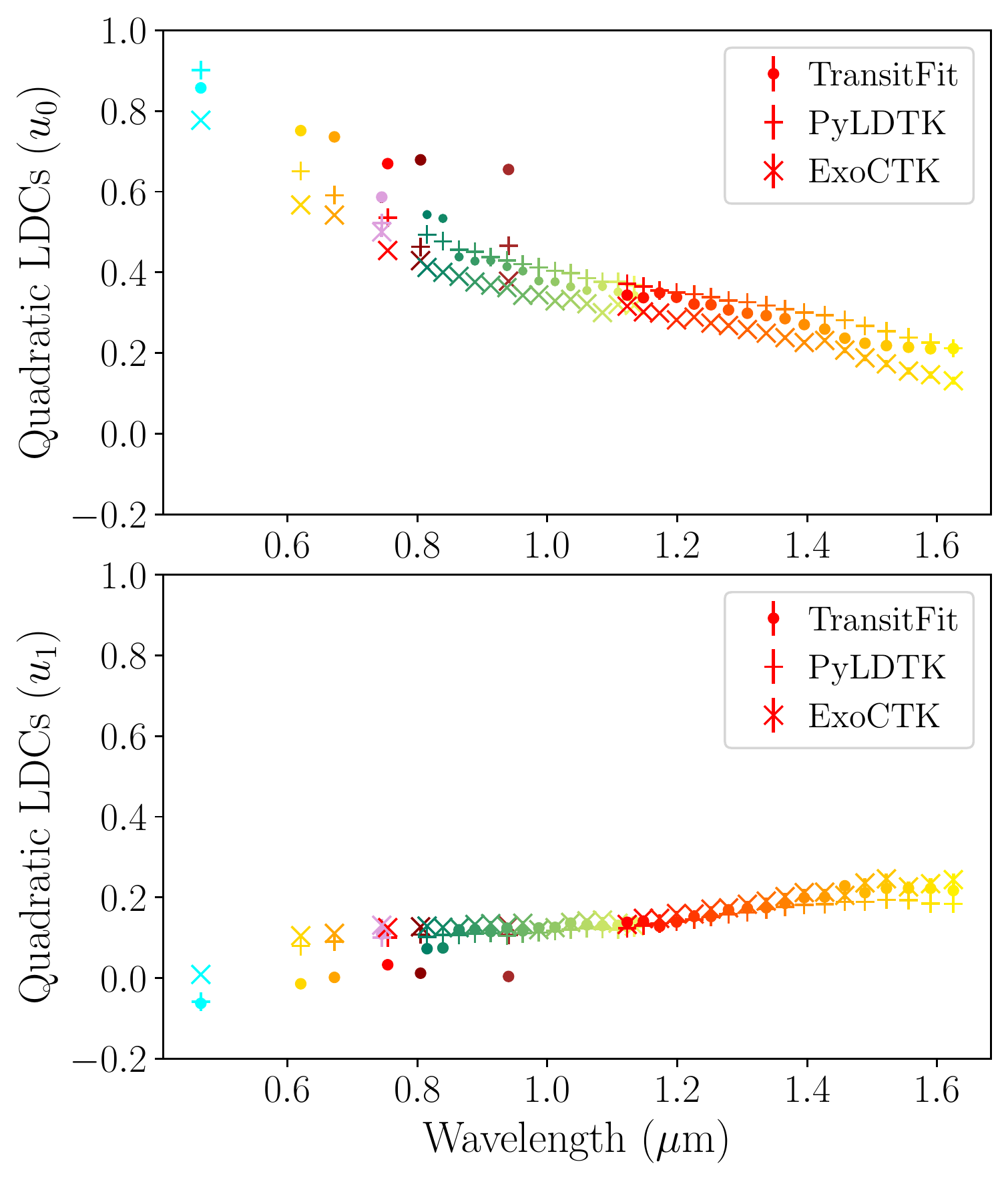} 
    \caption{The limb darkening coefficients were calculated from \texttt{TransitFit}, \texttt{LDTk} and ExoCTK. The colors of the bandpass filters are the same as those in \Cref{fig:Allan-TNT,fig:Allan-HST}.} 
   \label{fig:limb-dark}
\end{figure*}

\begin{figure*}[htb]
\centering
    \includegraphics[width=0.43\textwidth]{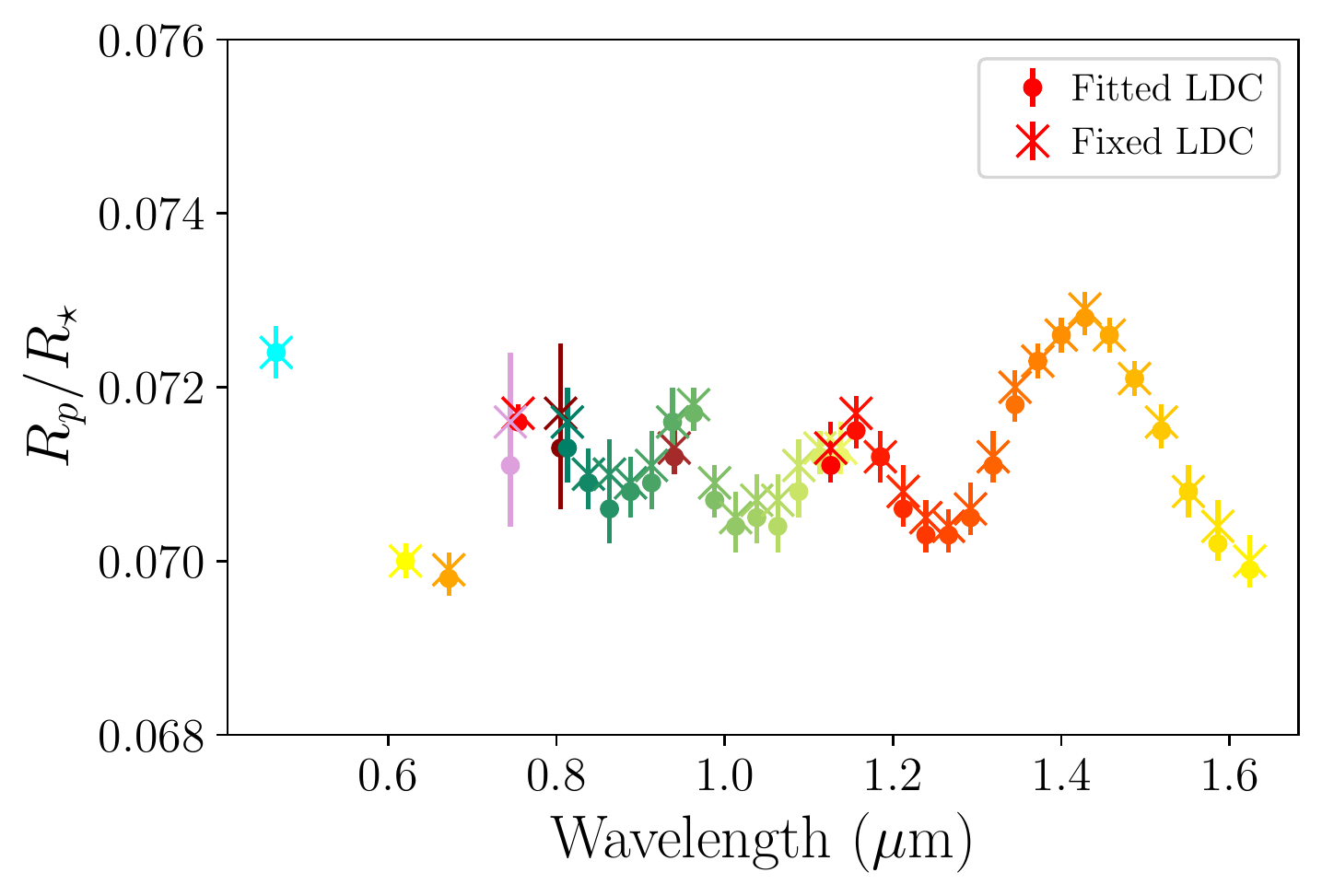} 
    \caption{The planetary radii calculated from both models with fitted LDCs and fixed LDCs. The colors of the bandpass filters are the same as those in \Cref{fig:Allan-TNT,fig:Allan-HST}.} 
   \label{fig:radii}
\end{figure*}

To validate our fitted limb darkening coefficient (LDC) values, we compare them against those acquired through the \texttt{LDTk} and the Exoplanet Characterization ToolKit \citep[ExoCTK,\footnote{ExoCTK limb darkening calculator: \texttt{https://exoctk.stsci.edu/limb\_darkening}}][]{ExoCTK}. Our fitted LDC values, the LDTk and ExoCTK limb darkening coefficients are plotted in Figure \ref{fig:limb-dark}. Although there is no overlap between the limb darkening values obtained from \texttt{TransitFit} and those from \texttt{LDTk} or ExoCTK, particularly within broad-band optical filters, the consistent trends are still apparent across all sources. To demonstrate that the discrepancy in limb darkening coefficients doesn't significantly impact the determination of planetary system parameters, especially the planetary radii, we conducted an analysis with the limb darkening coefficients fixed at the \texttt{LDTk} values. This analysis provided values of $i=87.92\pm0.06$ and $a/R_*=12.55\pm0.07$, aligning with the results from the fitting of LDCs. The mid-transit times fall within the 1-$\sigma$ range of the previous analysis. The planetary radii calculated with the fixed LDCs are presented in \Cref{tab:radius-transitDepth-limbdark}. These radii exhibit the same trend, with a slightly larger size that remains within the 1-$\sigma$ range of the analysis involving fitting LDCs, as shown in Figure \ref{fig:radii}. Therefore, we can confirm that these discrepancies do not effect the TTV and atmospheric analyses and use the data obtained from the fitted limb darkening coefficients in this work.

\begin{table*}
\begin{center}
\caption{The initial parameters and priors used to model the planetary parameters modeling with \texttt{TransitFit}.}
\label{tab:initialpara}          
\small\addtolength{\tabcolsep}{-2pt}
\begin{tabular}{lcc}
\toprule
Parameter  &  Priors &  Prior distribution \\
\hline
$P$ (days)         & 4.234516       &  Fixed      \\
${T}_{0}$ (BJD)    & 2455304.65122 $\pm$ 0.01 &  Gaussian  \\
$i$ (deg)          & 88.0 $\pm$ 0.5            &  Gaussian  \\
{$a/R_\ast$}       & 13 $\pm$ 1     &   Gaussian  \\
{$R_p$/$R_\ast$}   & (0.06, 0.08)   &   Uniform   \\
$e$                & 0              &   Fixed     \\
T$_{*}$ (K)        & $4700$         &   Fixed     \\
Z$_{*}$            & $-0.06$        &   Fixed     \\
$\textup{log}~(g_*)$ & $4.5$        &   Fixed     \\
\hline
\end{tabular}
\end{center}
{\textbf{Note} The priors of $P$, ${T}_{0}$, $i$ and {$a/R_\ast$} are set to the values in \citet{hartman2011}.}
\end{table*}

\begin{table*}
\begin{center}
\caption{The physical parameters of HAT-P-26 b from the \texttt{TransitFit} package and values from the literature.}
\label{tab:outpara}          
\small\addtolength{\tabcolsep}{-2pt}
\begin{tabular}{lccccc}
\toprule
\multirow{2}{*}{Parameter} & \multirow{2}{*}{\citet{hartman2011}}  & \multirow{2}{*}{\citet{steven2017}} & \multirow{2}{*}{\citet{von2019}} & \multicolumn{2}{c}{This work}  \\
\cline{5-6}
& & & & All LCs & Full LCs \\
\hline
$P$ (days)    &  4.234516 $\pm$ 2 $\times$ 10$^{-5}$   & 4.2345023 $\pm$ 7 $\times$ 10$^{-7}$  & 4.23450236 ± 3 $\times$ 10$^{-8}$ & \multicolumn{2}{c}{4.234516$^*$}\\
$i$ (deg)        &   $88.6^{+0.5}_{-0.9}$  &  87.3 $\pm$ 0.4   & 87.31 $\pm$ 0.09  & 87.82 $\pm$ 0.05 & 87.72 $\pm$ 0.05 \\
{$a/R_\ast$}     &   13.06  $\pm$	0.83    &  11.8 $\pm$ 0.6   & 12.05 $\pm$ 0.13  &  12.51 $\pm$ 0.07 & 12.53 $\pm$ 0.05 \\
\hline
\end{tabular}
\end{center}
{\textbf{Note} $^*$Value used is adopted from \citet{hartman2011}.}
\end{table*}

\begin{table*}
\begin{center}
\caption {HAT-P-26 b's mid-transit times ($T_{m}$) and timing residuals ($O-C$). $(O-C)_{33}$ are calculated from \Cref{eq:linear1}, which considers 33 mid-transit times modeled with the \texttt{TransitFit}. $(O-C)_{39}$ are calculated from \Cref{eq:linear2}, which include six mid-transit times from the literature, which have not published their raw light curves. Epoch~=~0 is the transit on 2010 April 18.}
\label{tab:midtransit}
\begin{tabular}{lcccc}
\toprule
\multirow{2}{*}{Epoch} & $T_{m} +2450000$ & $(O-C)_{33}$ &  $(O-C)_{39}$   & \multirow{2}{*}{Ref} \\
      & (BJD$_{TDB}$)    & (days) & (days)   &    \\
\hline
-105 &	4860.02786	$\pm$	0.00147$^\star$	&	-	&	-0.00146	&	(a)	\\
-24	&	5203.02521	$\pm$	0.00031	&	0.00116	&	0.00118	&	(a)	\\
-4	&	5287.71490	$\pm$	0.00050	&	0.00080	&	0.00082	&	(a)	\\
-3	&	5291.94879	$\pm$	0.00019	&	0.00019	&	0.00021	&	(a)	\\
0	&	5304.65218	$\pm$	0.00003$^\star$	& -	 &	0.00009	&	(b)	\\
5	&	5325.82444	$\pm$	0.00015	&	-0.00019	&	-0.00016	&	(a)	\\
9	&	5342.76192	$\pm$	0.00025	&	-0.00072	&	-0.00070	&	(a)	\\
260	&	6405.62370	$\pm$	0.0009$^\star$	&	- 	&	0.00094	&	(b)	\\
293	&	6545.36220	$\pm$	0.0003$^\star$	&	-	&	0.00085	&	(b)	\\
421	&	7087.37845	$\pm$	0.00032	&	0.00074	&	0.00077	&	(f)	\\
425	&	7104.31554	$\pm$	0.00023	&	-0.00018	&	-0.00014	&	(f)	\\
427	&	7112.78490	$\pm$	0.00057	&	0.00017	&	0.00021	&	(d)	\\
430	&	7125.48903	$\pm$	0.00071	&	0.00079	&	0.00083	&	(d)	\\
431	&	7129.72198	$\pm$	0.00090	&	-0.00076	&	-0.00073	&	(d)	\\
431	&	7129.72248	$\pm$	0.00017$^\star$	&	- 	&	-0.00022	&	(b)	\\
439	&	7163.59815	$\pm$	0.00054	&	-0.00061	&	-0.00057	&	(d)	\\
443	&	7180.53670	$\pm$	0.00041	&	-0.00007	&	-0.00004	&	(d)	\\
447	&	7197.47394	$\pm$	0.00024	&	-0.00084	&	-0.00081	&	(d)	\\
498	&	7413.43284	$\pm$	0.00017$^\star$ &	-	&	-0.00154	&	(c)	\\
502	&	7430.37175	$\pm$	0.00025	&	-0.00068	&	-0.00064	&	(f)	\\
509	&	7460.01268	$\pm$	0.00005	&	-0.00126	&	-0.00122	&	(c)	\\
521	&	7510.82651	$\pm$	0.00005	&	-0.00146	&	-0.00143	&	(c)	\\
524	&	7523.53019	$\pm$	0.00099	&	-0.00129	&	-0.00125	&	(d)	\\
546	&	7616.68959	$\pm$	0.00007	&	-0.00095	&	-0.00091	&	(c)	\\
596	&	7828.41858	$\pm$	0.00093	&	0.00291	&	0.00295	&	(f)	\\
610	&	7887.70232	$\pm$	0.00420	&	0.00362	&	0.00366	&	(d)	\\
614	&	7904.63921	$\pm$	0.00079	&	0.00249	&	0.00253	&	(d)	\\
618	&	7921.57729	$\pm$	0.00045	&	0.00257	&	0.00260	&	(d)	\\
685	&	8205.28614	$\pm$	0.00047	&	-0.00026	&	-0.00022	&	(f)	\\
689	&	8222.22421	$\pm$	0.00022	&	-0.00020	&	-0.00016	&	(f)	\\
690	&	8226.45946	$\pm$	0.00065	&	0.00055	&	0.00059	&	(d)	\\
766	&	8548.28051	$\pm$	0.00017	&	-0.00060	&	-0.00056	&	(f)	\\
778	&	8599.09456	$\pm$	0.00038	&	-0.00058	&	-0.00054	&	(f)	\\
1029 &	9661.95548	$\pm$	0.00078	&	0.00018	&	0.00023	&	(f)	\\
1031 &	9670.42414	$\pm$	0.00076	&	-0.00016	&	-0.00011	&	(e)	\\
1032 &	9674.65912	$\pm$	0.00078	&	0.00032	&	0.00037	&	(e)	\\
1035 &	9687.36206	$\pm$	0.00075	&	-0.00026	&	-0.00020	&	(e)	\\
1041 &	9712.77013	$\pm$	0.00171	&	0.00080	&	0.00086	&	(f)	\\
1045 &	9729.70826	$\pm$	0.00091	&	0.00092	&	0.00097	&	(f)	\\

\hline
\end{tabular}\\
\end{center}
{\textbf{Notes} Data Source: (a) \citet{hartman2011} (b) \citet{steven2017} (c) \citet{wak2017}, (d) \citet{von2019}, (e) TESS and (f) this study. \\ 
$^\star$ : $T_m$ adopted from the literature. \\}
\end{table*}

\begin{table*}
\begin{center}
\caption{The planet-to-star radius ratio ($R_p$/$R_\ast$) for both models with fitted LDCs and fixed LDCs, and quadratic LDCs of HAT-P-26~b in different filters, as obtained by \texttt{TransitFit} using \texttt{coupled} fitting of LDCs.}
\label{tab:radius-transitDepth-limbdark}          
\begin{tabular}{ccccccc}
\toprule
\multirow{2}{*}{Filter} & Mid-wavelength & Bandwidth & \multicolumn{2}{c}{$R_p$/$R_\ast$} & \multirow{2}{*}{$u_0$} & \multirow{2}{*}{$u_1$} \\ 
 & ($\mu$m) & ($\mu$m) &	Fitted LDC & Fixed LDC &	 &  \\ 
\hline				
$g'$-band		&	0.467	&	0.139	&	0.0724	$\pm$	0.0003	&	0.0724	$\pm$	0.0003	&	0.857	$\pm$	0.013	&	-0.063	$\pm$	0.012	\\
$r'$-band		&	0.621	&	0.124	&	0.0700	$\pm$	0.0002	&	0.0700	$\pm$	0.0002	&	0.751	$\pm$	0.012	&	-0.014	$\pm$	0.012	\\
$i'$-band		&	0.754	&	0.130	&	0.0716	$\pm$	0.0001	&	0.0717	$\pm$	0.0001	&	0.669	$\pm$	0.012	&	0.033	$\pm$	0.011	\\
$z'$-band		&	0.940	&	0.256	&	0.0712	$\pm$	0.0002	&	0.0713	$\pm$	0.0003	&	0.655	$\pm$	0.013	&	0.004	$\pm$	0.012	\\
$R$-band		&	0.672	&	0.107	&	0.0698	$\pm$	0.0002	&	0.0699	$\pm$	0.0002	&	0.736	$\pm$	0.013	&	0.002	$\pm$	0.012	\\
$I$-band		&	0.805	&	0.289	&	0.0713	$\pm$	0.0007	&	0.0717	$\pm$	0.0008	&	0.679	$\pm$	0.013	&	0.012	$\pm$	0.012	\\
TESS		&	0.745	&	0.400	&	0.0711	$\pm$	0.0007	&	0.0716	$\pm$	0.0008	&	0.587	$\pm$	0.014	&	0.109	$\pm$	0.013	\\
HST/WFC3	G102	&	0.813	&	0.025	&	0.0713	$\pm$	0.0004	&	0.0716	$\pm$	0.0004	&	0.543	$\pm$	0.007	&	0.073	$\pm$	0.007	\\
HST/WFC3	G102	&	0.838	&	0.025	&	0.0709	$\pm$	0.0003	&	0.0710	$\pm$	0.0003	&	0.534	$\pm$	0.007	&	0.075	$\pm$	0.007	\\
HST/WFC3	G102	&	0.863	&	0.025	&	0.0706	$\pm$	0.0004	&	0.0710	$\pm$	0.0004	&	0.438	$\pm$	0.009	&	0.120	$\pm$	0.009	\\
HST/WFC3	G102	&	0.888	&	0.025	&	0.0708	$\pm$	0.0003	&	0.0709	$\pm$	0.0003	&	0.428	$\pm$	0.008	&	0.120	$\pm$	0.008	\\
HST/WFC3	G102	&	0.913	&	0.025	&	0.0709	$\pm$	0.0003	&	0.0711	$\pm$	0.0004	&	0.429	$\pm$	0.009	&	0.115	$\pm$	0.008	\\
HST/WFC3	G102	&	0.938	&	0.025	&	0.0716	$\pm$	0.0003	&	0.0716	$\pm$	0.0004	&	0.414	$\pm$	0.008	&	0.124	$\pm$	0.008	\\
HST/WFC3	G102	&	0.963	&	0.025	&	0.0717	$\pm$	0.0002	&	0.0718	$\pm$	0.0002	&	0.403	$\pm$	0.005	&	0.117	$\pm$	0.005	\\
HST/WFC3	G102	&	0.988	&	0.025	&	0.0707	$\pm$	0.0002	&	0.0709	$\pm$	0.0002	&	0.379	$\pm$	0.005	&	0.124	$\pm$	0.006	\\
HST/WFC3	G102	&	1.013	&	0.025	&	0.0704	$\pm$	0.0003	&	0.0705	$\pm$	0.0003	&	0.377	$\pm$	0.007	&	0.126	$\pm$	0.007	\\
HST/WFC3	G102	&	1.038	&	0.025	&	0.0705	$\pm$	0.0003	&	0.0707	$\pm$	0.0003	&	0.364	$\pm$	0.007	&	0.136	$\pm$	0.008	\\
HST/WFC3	G102	&	1.063	&	0.025	&	0.0704	$\pm$	0.0003	&	0.0707	$\pm$	0.0003	&	0.355	$\pm$	0.007	&	0.132	$\pm$	0.007	\\
HST/WFC3	G102	&	1.088	&	0.025	&	0.0708	$\pm$	0.0003	&	0.0711	$\pm$	0.0003	&	0.365	$\pm$	0.007	&	0.130	$\pm$	0.007	\\
HST/WFC3	G102	&	1.113	&	0.025	&	0.0712	$\pm$	0.0002	&	0.0713	$\pm$	0.0002	&	0.351	$\pm$	0.005	&	0.131	$\pm$	0.005	\\
HST/WFC3	G102	&	1.138	&	0.025	&	0.0712	$\pm$	0.0002	&	0.0713	$\pm$	0.0003	&	0.343	$\pm$	0.005	&	0.140	$\pm$	0.005	\\
HST/WFC3	G141	&	1.126	&	0.031	&	0.0711	$\pm$	0.0002	&	0.0713	$\pm$	0.0003	&	0.343	$\pm$	0.006	&	0.138	$\pm$	0.007	\\
HST/WFC3	G141	&	1.156	&	0.029	&	0.0715	$\pm$	0.0002	&	0.0717	$\pm$	0.0002	&	0.337	$\pm$	0.006	&	0.141	$\pm$	0.007	\\
HST/WFC3	G141	&	1.185	&	0.028	&	0.0712	$\pm$	0.0002	&	0.0712	$\pm$	0.0003	&	0.349	$\pm$	0.006	&	0.128	$\pm$	0.007	\\
HST/WFC3	G141	&	1.212	&	0.027	&	0.0706	$\pm$	0.0002	&	0.0708	$\pm$	0.0003	&	0.338	$\pm$	0.006	&	0.139	$\pm$	0.007	\\
HST/WFC3	G141	&	1.239	&	0.027	&	0.0703	$\pm$	0.0002	&	0.0705	$\pm$	0.0002	&	0.321	$\pm$	0.005	&	0.153	$\pm$	0.005	\\
HST/WFC3	G141	&	1.266	&	0.027	&	0.0703	$\pm$	0.0002	&	0.0704	$\pm$	0.0002	&	0.320	$\pm$	0.005	&	0.153	$\pm$	0.005	\\
HST/WFC3	G141	&	1.292	&	0.027	&	0.0705	$\pm$	0.0002	&	0.0706	$\pm$	0.0003	&	0.307	$\pm$	0.007	&	0.169	$\pm$	0.008	\\
HST/WFC3	G141	&	1.319	&	0.026	&	0.0711	$\pm$	0.0002	&	0.0712	$\pm$	0.0003	&	0.299	$\pm$	0.007	&	0.173	$\pm$	0.008	\\
HST/WFC3	G141	&	1.345	&	0.027	&	0.0718	$\pm$	0.0002	&	0.0720	$\pm$	0.0002	&	0.293	$\pm$	0.007	&	0.174	$\pm$	0.008	\\
HST/WFC3	G141	&	1.372	&	0.027	&	0.0723	$\pm$	0.0002	&	0.0723	$\pm$	0.0002	&	0.285	$\pm$	0.007	&	0.185	$\pm$	0.008	\\
HST/WFC3	G141	&	1.400	&	0.028	&	0.0726	$\pm$	0.0002	&	0.0726	$\pm$	0.0002	&	0.270	$\pm$	0.005	&	0.199	$\pm$	0.007	\\
HST/WFC3	G141	&	1.428	&	0.029	&	0.0728	$\pm$	0.0002	&	0.0729	$\pm$	0.0002	&	0.259	$\pm$	0.005	&	0.201	$\pm$	0.007	\\
HST/WFC3	G141	&	1.457	&	0.029	&	0.0726	$\pm$	0.0002	&	0.0726	$\pm$	0.0002	&	0.237	$\pm$	0.007	&	0.229	$\pm$	0.010	\\
HST/WFC3	G141	&	1.487	&	0.031	&	0.0721	$\pm$	0.0002	&	0.0721	$\pm$	0.0002	&	0.224	$\pm$	0.008	&	0.212	$\pm$	0.010	\\
HST/WFC3	G141	&	1.519	&	0.032	&	0.0715	$\pm$	0.0002	&	0.0716	$\pm$	0.0002	&	0.218	$\pm$	0.008	&	0.223	$\pm$	0.011	\\
HST/WFC3	G141	&	1.551	&	0.034	&	0.0708	$\pm$	0.0002	&	0.0708	$\pm$	0.0003	&	0.214	$\pm$	0.007	&	0.223	$\pm$	0.010	\\
HST/WFC3	G141	&	1.586	&	0.036	&	0.0702	$\pm$	0.0002	&	0.0704	$\pm$	0.0003	&	0.211	$\pm$	0.008	&	0.222	$\pm$	0.011	\\
HST/WFC3	G141	&	1.624	&	0.039	&	0.0699	$\pm$	0.0002	&	0.0700	$\pm$	0.0003	&	0.211	$\pm$	0.008	&	0.217	$\pm$	0.010	\\
\hline
\end{tabular}
\end{center}
\end{table*}

\section{Transit-timing analysis}
\label{sec:ttv}
\subsection{A Refined Ephemeris}
The mid-transit times of 33 epochs obtained from \texttt{TransitFit}, and listed in \Cref{tab:midtransit}, are considered for our timing analysis. The mid-transit times were fitted by a linear ephemeris model, using a constant-period as:  
\begin{equation}
T^{c}_{m}(E) =  T_{0,l} + P_{l} \times E \ ,
\end{equation}
where $T_{0,l}$ and $P_{l}$ are the reference time and the orbital period of the linear ephemeris model, respectively, $E$ is the epoch number and $E = 0$ represents the transit on 2010 April 18. $T^{c}_{m}(E)$ is the calculated mid-transit time at a given epoch $E$.

To find the best-fit parameters from the model, used \texttt{emcee} \citep{foreman2013} to perform a Markov Chain Monte Carlo (MCMC) fit with 50 chains and $10^{5}$ MCMC steps. The new linear ephemeris was defined as:
\begin{equation}
\label{eq:linear1}
T^{c}_{m}(E) =  2455304.65211^{+0.00036}_{-0.00035} +  4.234503^{+0.000001}_{-0.000001} E\ .
\end{equation}
The reduced chi-square of the linear fit is ${\chi}^{2}_{red}$ = 55 with 31 degrees of freedom. The Bayesian Information Criterion, $BIC = \chi^{2} + k \ln n = 1698$, where $k$ is the number of free parameters, and $n$ is the number of data points. A corner plot indicating the MCMC posterior probability distribution of the parameters is shown in \Cref{fig:linearMCMC_TransitFit}. The obtained period from the $O-C$ is consistent with the period provided by \citet{steven2017,von2019}. However, the value differs from our prior period in the \texttt{TransitFit}, which adopt from \citet{hartman2011}, by $sim$1 s. The difference does not affect our fitted timing as we used the \texttt{allow\_TTV} function in the \texttt{TransitFit}. For the fitted physical parameters, the effects of the different periods are small and negligible. Using the new ephemeris, we constructed an $O-C$ diagram (\Cref{fig:oc-gls_f0.004}b), which shows the residual difference between the timing data and \Cref{eq:linear1}. 

In addition to the mid-transit times obtained from \texttt{TransitFit}, there are six transits whose light curves are not publicly available for refitting, so only their published transit times can be used (as listed in \Cref{tab:midtransit}). When added to the 33 transit times fitted with \texttt{TransitFit}, the combined 39 mid-transit times were linearly fitted using the same MCMC procedure, resulting in the following revised linear ephemeris
\begin{equation}
\label{eq:linear2}
T^{c}_{m}(E) = 2455304.65209^{+0.00030}_{-0.00030} + 4.234503^{+0.000001}_{-0.000001} E\ .
\end{equation}
The MCMC posterior probability distribution for these 39 epochs is shown in \Cref{fig:linearMCMC_TransitFitLit}. The best-fitting model shows ${\chi}^{2}_{red}$ = 46 with 37 degrees of freedom and BIC = 1713. Using the ephemeris from this linear fitting, another $O-C$ diagram was constructed, shown in \Cref{fig:oc-gls_multi-f}.

\subsection{The Frequency Analysis of TTVs}
\label{subsec:gls}
The previous TTV analysis of HAT-P-26~b by \citet{von2019} found cyclic variation with a period of $\sim270$ epochs. In this work, we re-investigate the TTVs using the timing from our refitting result in \Cref{tab:midtransit}. The Generalized Lomb-Scargle periodogram \citep[GLS,][]{zech2009} from the \texttt{PyAstronomy}\footnote{PyAstronomy: \texttt{https://github.com/sczesla/PyAstronomy}} routines \citep{zesla2019} was used to search for periodicity in the timing-residual data.

\begin{figure*}[htb]
\centering
  \begin{tabular}{cc}
    \includegraphics[width=0.49\textwidth,page=1]{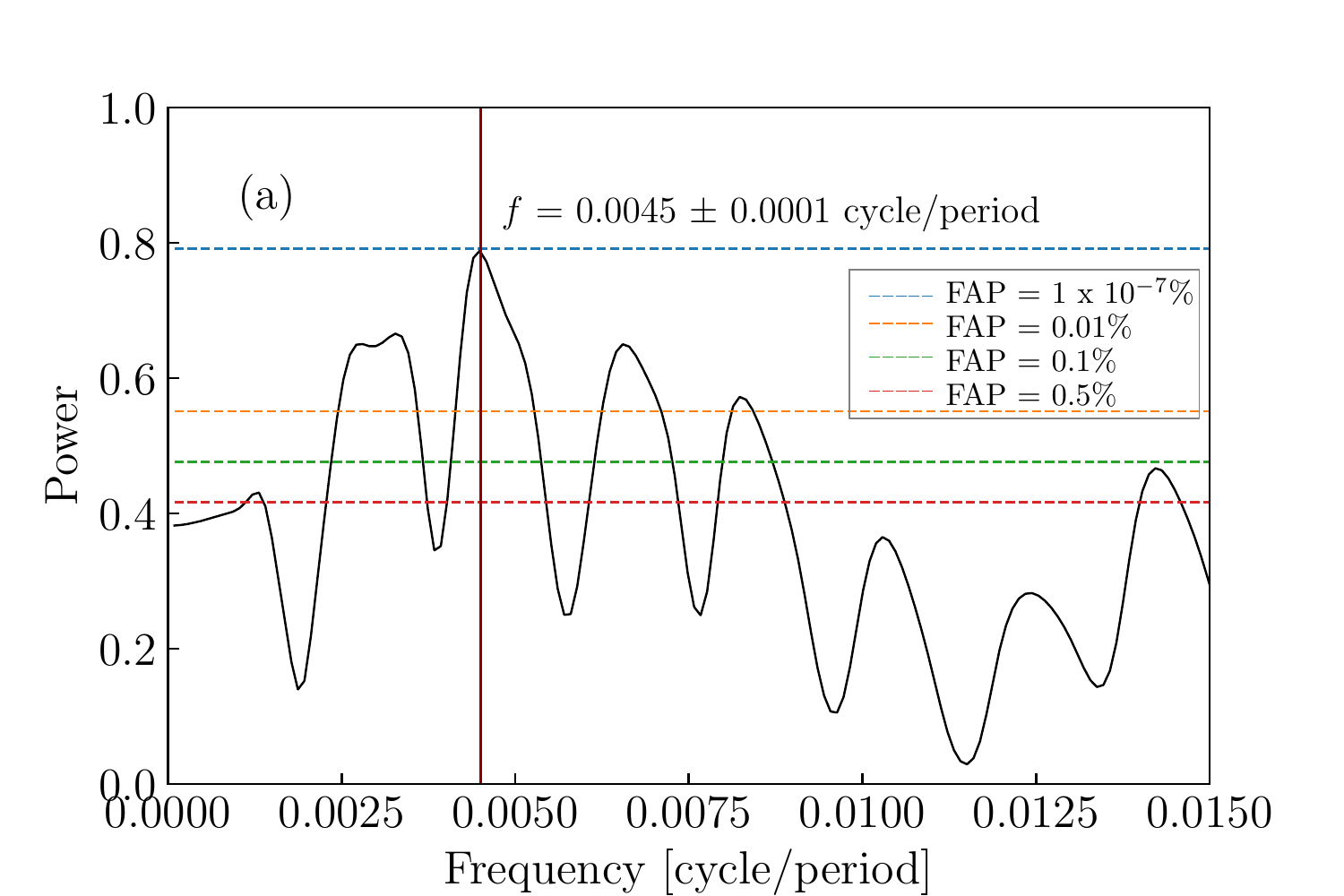} \\
    \includegraphics[width=0.5\textwidth,page=1, trim ={0 0 0 0.3cm}]{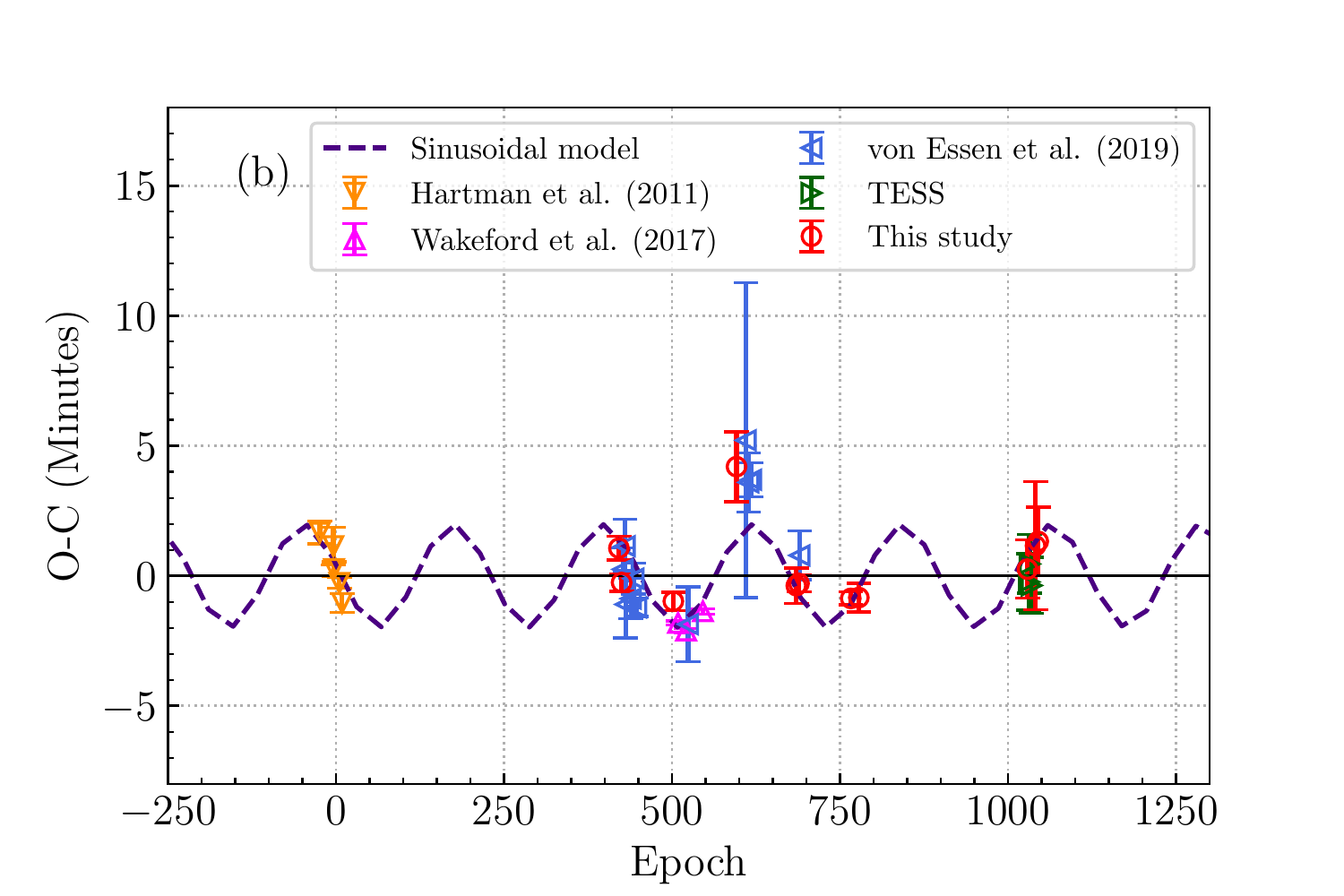} &
    \end{tabular}
    \caption{(a) GLS periodogram for timing residuals of 33 mid-transit times obtained from the \texttt{TransitFit}. The dashed lines indicate the FAP levels. (b) $O-C$ diagram and the best fit of sinusoidal variability from the frequency of the highest power peak, FAP = 1 $\times$ 10$^{-7}$ \% (purple dashed line).}
   \label{fig:oc-gls_f0.004}
\end{figure*}

Firstly, we performed a GLS analysis on our 33 refitted timing residuals based on \Cref{eq:linear1}. The result is shown as a periodogram in \Cref{fig:oc-gls_f0.004}a. In this periodogram, the highest-power peak has a strength of 0.7882 at a frequency of 0.0045 $\pm$ 0.0001 cycles/period ($\simeq222$ epochs) with a false alarm probability (FAP) of 1 $\times$ 10$^{-7}$ \%.

The frequency of the highest-power peak is assumed to be the frequency of the cyclic TTV of the system. In order to find the amplitude of the cyclic variation, the same procedure as described in \citet{von2019} is used. The timing residuals were fitted through a fitting function: 
\begin{equation}
\label{eq:frequency}
TTV(E) = A_{TTV}\sin(2{\pi}fE - \phi) \  ,
\end{equation}
where $A_{TTV}$ is the amplitude (in minutes) of the timing perturbation, $f$ is the frequency on the highest peak of the power periodogram, and $\phi$ is the orbital phase at $E=0$. From the fitting, an amplitude of $A_{TTV}$ = 1.98 $\pm$ 0.05 minutes and an initial orbital phase of $\phi$ = -0.22 $\pm$ 0.04 is obtained. The best-fitting model provides ${\chi}^{2}_{\rm red}$ = 4.2 and BIC = 136.2. The timing residuals with the best fit of sinusoidal variability are plotted in \Cref{fig:oc-gls_f0.004}b. This period is much shorter than the period obtained by \citet{von2019}. 

The difference in the TTV period might be caused by differences in our datasets. There are six transit times in \citet{von2019}'s analysis (one transit time from \citet{hartman2011}, four transit times from \citet{steven2017} and one transit time from \citet{wak2017}) which have not been used in this work, as their raw light curves have yet to be published. In order to answer whether these six transit times affect the TTV periodicity, we also perform the GLS analysis on the combined set of 39 epochs, using the ephemeris of \Cref{eq:linear2}. 

\begin{figure*}[htb]
\centering
  \begin{tabular}{cc}
    \includegraphics[width=0.5\textwidth,page=1]{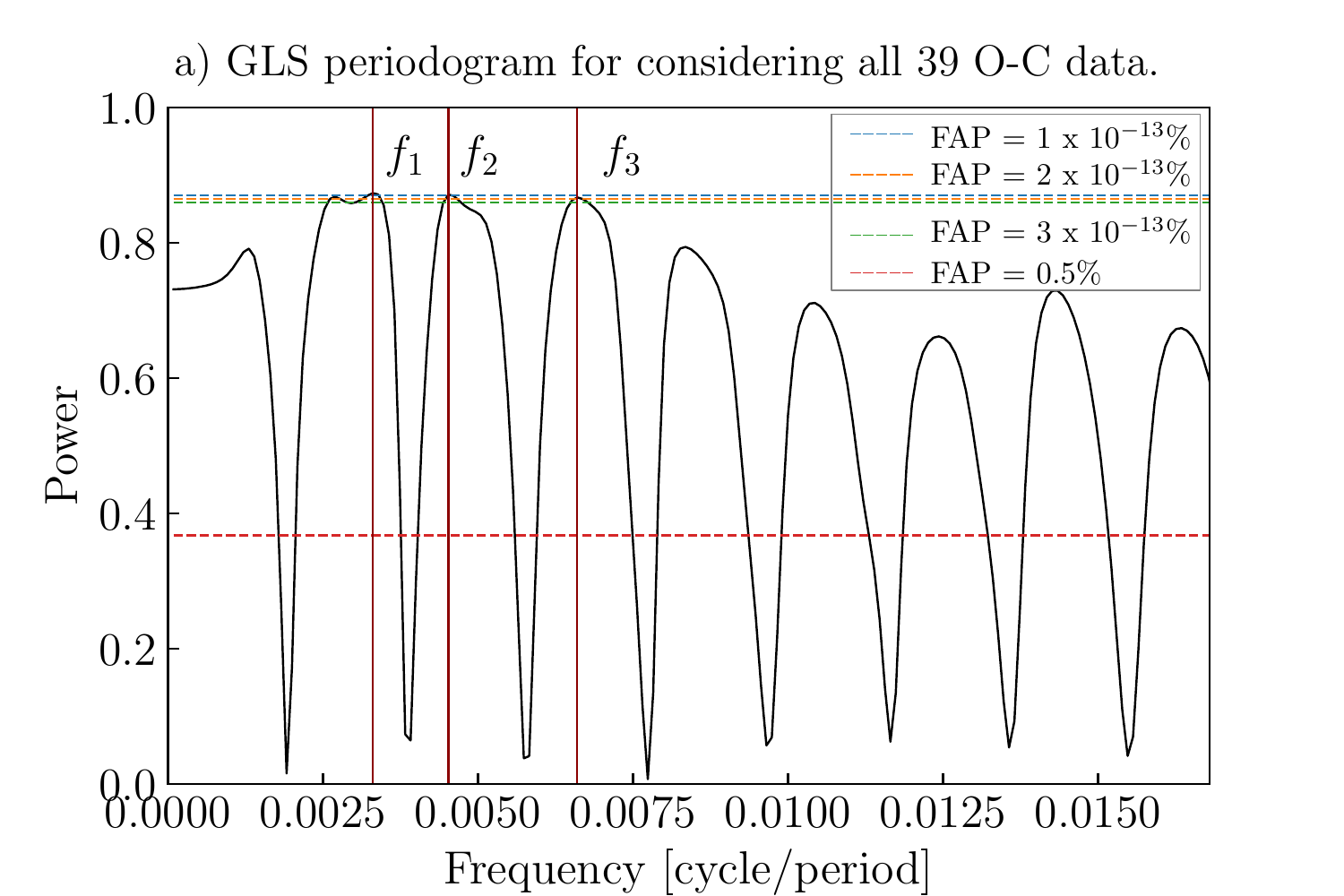} \\
    \includegraphics[width=0.49\textwidth,page=1,trim ={0 0 0 0.2cm}]{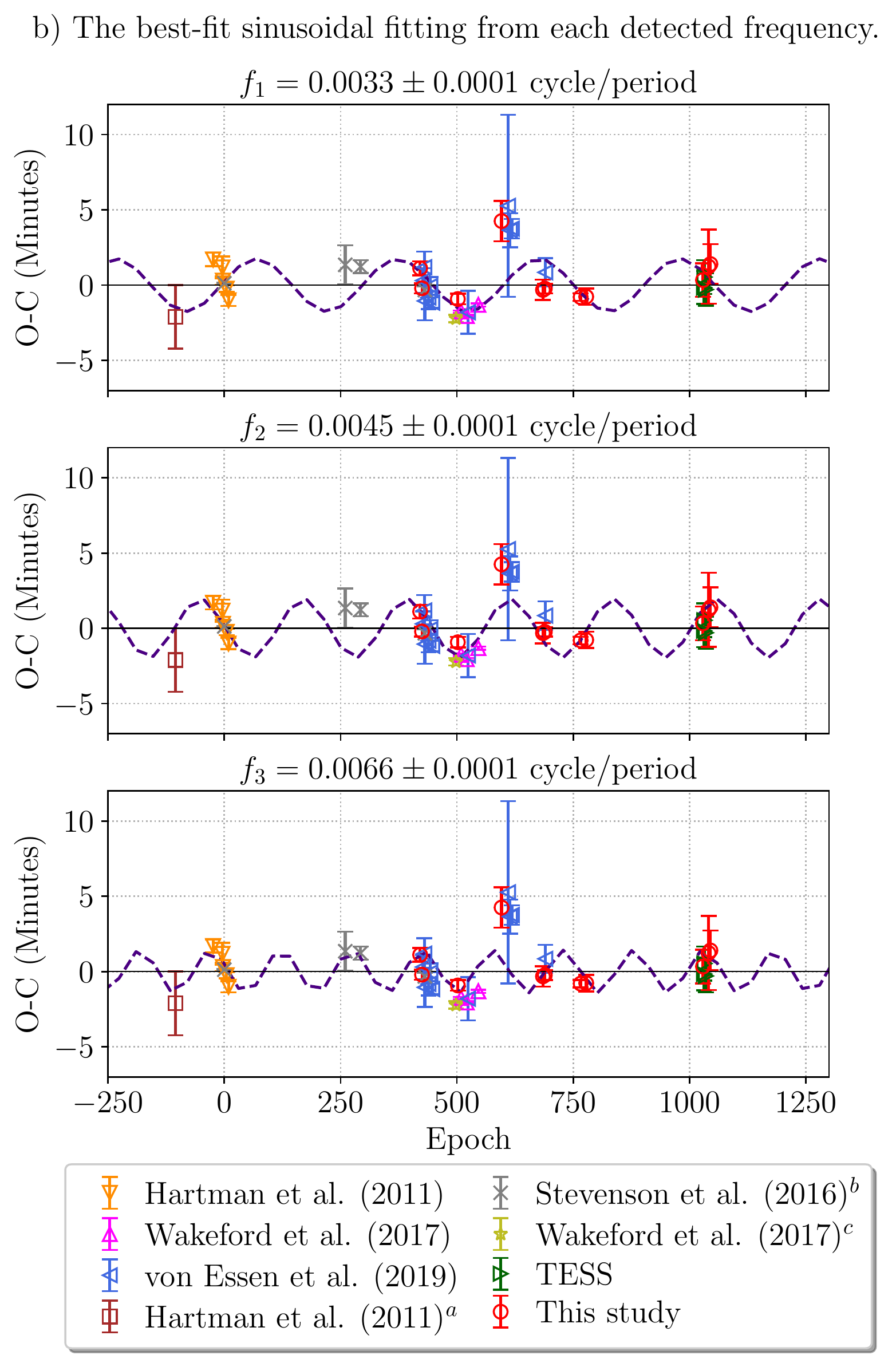} &
    \end{tabular}
    \caption{(a) GLS periodogram of timing residuals of all 39 mid-transit times, showing three peaks with FAP $<$ 3 $\times$ 10$^{-13}$$\%$. The dashed lines indicate the FAP levels. (b) $O-C$ diagram and the best-fit sinusoid for the detected frequencies $f_{1}$ (top), $f_{2}$ (middle), and $f_{3}$ (bottom; purple dashed line).}
\raggedright
    {\textbf{Notes}: $^a$ : Mid-transit times ($T_m$) are adopted from the transit light curve observed by HATNet field 376 \citep{hartman2011}, \\$^b$ : $T_m$ adopted from \citet{steven2017}, \\$^c$ : $T_m$ adopted from the transit light curves observed by HST/STIS \citep{wak2017}.}
    \label{fig:oc-gls_multi-f}
\end{figure*}

The GLS analysis for these 39 epochs detects three periodicity peaks with FAP 3 $\times$ 10$^{-13}$$\%$, shown in \Cref{fig:oc-gls_multi-f}a. The three corresponding best-fit sinusoidal functions are shown in \Cref{tab:multi-f}. The timing residuals with the best-fit sinusoidal variability for each power peak detection are plotted in \Cref{fig:oc-gls_multi-f}b. From these three power peaks, there is a peak with a frequency of 0.0045 $\pm$ 0.0001 cycle$/$period, which has a frequency similar to the frequency of the power peak of the 33 \texttt{TransitFit} refitted timing. While the other two peak frequencies, $f_1$ and $f_3$, could be harmonics of this frequency ($f_2$). Therefore, HAT-P-26~b timing is consistent with a sinusoidal variation with a frequency of 0.0045 $\pm$ 0.0001 cycle$/$period.

There are many possible causes of the TTV signal. For example, stellar activity \citet{Rabus2009,Barros2013} or gravitational interaction from an additional planet in the system. The variations due to the stellar activity are likely to be ruled out as \citet{von2019} show that there is no spot modulation within the precision limit of the data within three years. We therefore instead consider the possibility of the presence of an additional planet.

Given the frequency of 0.0045 $\pm$ 0.0001 cycle$/$period, and the assumption of a co-planar orbit, we model an additional exoplanet with an orbital period near the first-order resonance of HAT-P-26~b. In case of a first-order mean-motion resonance, \textit{j:j-1}, \citet{lith2012} allows us to calculate the additional planet mass as:
\begin{equation}
    V = P\frac{\mu^{'}}{\pi j^{2/3} (j-1)^{1/3} \Delta} \left(-f-\frac{3}{2} \frac{Z^{*}_{free}}{\Delta}\right) \ ,
\label{eq:masspert}
\end{equation}
where $V$ is the amplitude of TTV (from our analysis, $V$ = 1.98 minutes), \textit{P} is the orbital period of HAT-P-26 b, $\mu^{'}$ is the outer-planet mass, $\Delta$ is the normalized distance to resonance, $f$ is the sum of the Laplace coefficients with order-unity values, and $Z^{*}_{free}$ is the dynamical quantity that controls the TTV signal. From the analysis, if an additional planet has 2:1 orbital resonance with HAT-P-26~b (i.e.$P \sim8.47$ days), we find that the mass of the additional planet could be around 0.02M$_{\textup{Jup}}$ (6.36M$_{\oplus}$).  

\begin{table*}
\begin{center}
\caption{The detected frequencies with the best-fit parameters of the sinusoidal functions, considering all 39 mid-transit times.}
\label{tab:multi-f}          
\small\addtolength{\tabcolsep}{-2pt}
\begin{tabular}{ccccccc}
\toprule
Frequencies & \multirow{2}{*}{Power} & \multirow{2}{*}{FAP}  &$A_{TTV}$ & \multirow{2}{*}{$\phi$} & \multirow{2}{*}{${\chi}^{2}_{red}$} & \multirow{2}{*}{BIC} \\
(cycle$/$period) &    &      &  (minutes) &  &  &  \\
\hline
$f_{1}$	= 0.0033 $\pm$ 0.0001 &	0.873 &	1 $\times$ 10$^{-13}$$\%$ &	1.75 $\pm$ 0.05	& -0.13	$\pm$ 0.02	&	6.69 & 254.72 \\
$f_{2}$	= 0.0045 $\pm$ 0.0001 &	0.871 &	2 $\times$ 10$^{-13}$$\%$ &	1.95 $\pm$ 0.05	& 3.23	$\pm$ 0.02	&	6.13 & 233.95	\\
$f_{3}$	= 0.0066 $\pm$ 0.0001 &	0.867 &	3 $\times$ 10$^{-13}$$\%$ &	1.42 $\pm$ 0.06	& -0.36	$\pm$ 0.03	&	22.65 &	845.39	\\
\hline
\end{tabular}
\end{center}
\end{table*}

\section{Atmospheric Modeling}
\label{sec:atmosphere}
\begin{figure*}[htb]
\centering
    \includegraphics[width=0.49\textwidth,page=1]{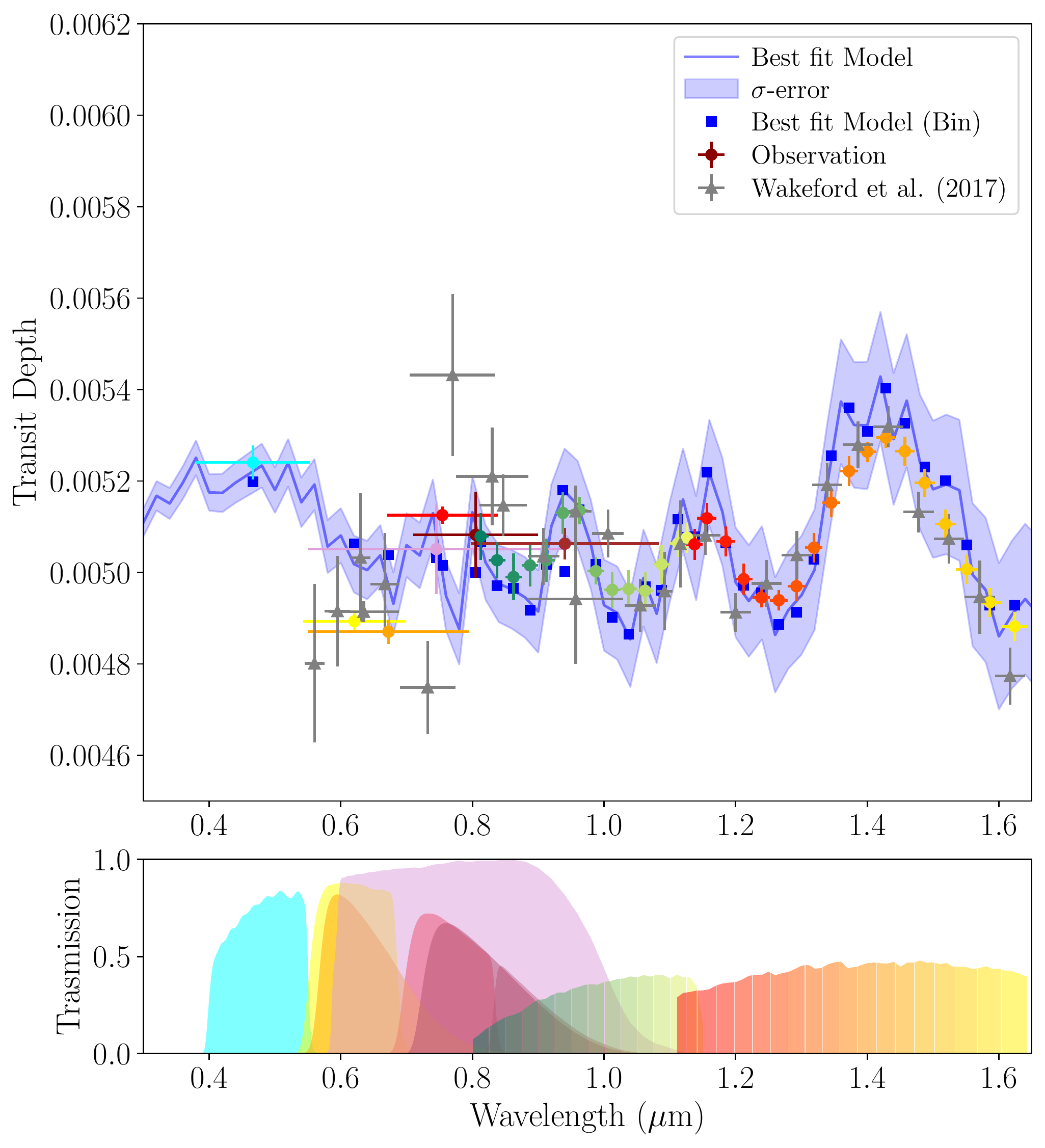} 
    
    \caption{Top panel: The best-fit transmission spectrum of HAT-P-26~b, calculated using only all ground-based and the HST/WFC3 data from \texttt{TransitFit}, with synthetic models generated by \texttt{TauREx 3} (blue solid line), with their 1$\sigma$ error (blue bands). The blue squares are the binned best-fit transmission spectra. The gray stars are the transmission spectra obtained by \citet{wak2017}. The observed data are binned using the bandpass in the bottom panel.}
    \label{fig:transmission_spectrum}
\end{figure*}

Previous studies of HAT-P-26~b via near-infrared transmission spectroscopy found a significant detection of H$_2$O in the atmosphere \citep{steven2017, wak2017,macdonald2019}. Optical analysis by \citet{macdonald2019} found evidence of metal hydrides, with three potential candidates identified as TiH, CrH, and ScH. The derived temperature from their study was $563^{+59}_{-55}$ K, with a temperature gradient of $\sim$80 K. To confirm the presence of metal hydrides in the optical and the H$_2$O at near-infrared wavelengths, we re-investigated the chemical composition of the HAT-P-26~b's atmosphere using the combined spectrophotometry from optical ground-based observations and the optical/near-infrared observations by HST. Our fitted $R_p/R_{\star}$ values using \texttt{TransitFit} are consistent with the values provided by \citet{wak2017} in both optical and near-infrared wavebands as shown in \Cref{fig:transmission_spectrum}.

Retrieval of the transmission spectrum was performed using the open-source atmospheric retrieval framework \citep[TauREx~3,\footnote{\texttt{TauREx~3}: \texttt{https://github.com/ucl-exoplanets/TauREx3\_public/}}][]{al-rafaie2021}  using the nested sampling routines from \texttt{MultiNest} \citep{feroz2009} with 1000 live points. The 38 transit depths from \Cref{tab:radius-transitDepth-limbdark} are used to retrieve planetary atmospheric compositions. The stellar parameters and the planet mass were adopted from \citet{hartman2011}. The stellar emission spectrum was simulated from a \texttt{PHOENIX} model \citep{husser2013} for a star of $T_* = 4700$ K. We adopted an isothermal temperature profile and a parallel plane atmosphere of 100 layers, with a pressure ranging from $10^{-1}$ to $10^{6}$ Pa with logarithmic spacing.

In keeping with \cite{macdonald2019}, we modelled the molecular opacities of metal hydrides, including TiH \citep{burrows2005}, CrH \citep{burrows2002} and ScH \citep{lodi2015}. We also added the presence of the following active trace gases: TiO \citep{mckemmish2019}, VO \citep{mckemmish2016}, K and Na \citep{allard2019}, MgH \citep{owen2022}, SiH \citep{yurchenko2018}, N$_{2}$ \citep{western2018}, O$_{2}$ \citep{somogyi2021} and H$_{2}$O \citep{polyansky2018}, and the inactive gases of He/H$_{2}$ \citep{abel2012}. The molecular line lists are taken from the ExoMol \citep{tennyson2016J}, HITRAN \citep{gordon2016}, and HITEMP \citep{rothman2014} databases. We also include collision-induced absorption between H$_2$ molecules \citep{abel2011,fletcher2018} and between H$_{2}$ and He \citep{abel2012} in the transmission spectrum model. A list of the parameters used in the \texttt{TauREx 3} retrieval is shown in \Cref{tab:atmosphere_para}.

\begin{figure*}[htb]
\centering
    \includegraphics[width=1\textwidth,page=1]{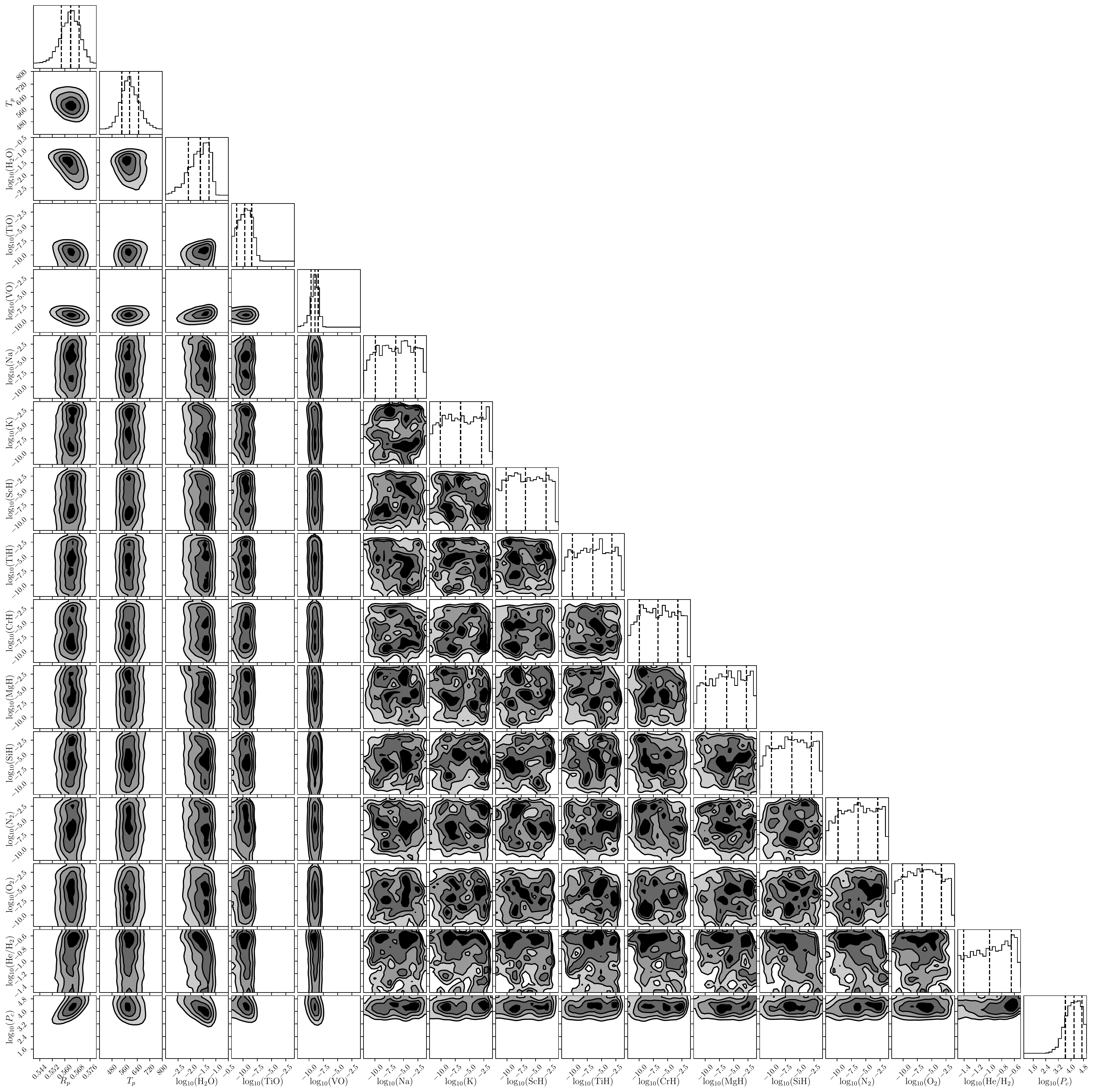} 
    \caption{Posterior probability distribution for our HAT-P-26~b atmosphric model, using nested sampling and the \texttt{TauREx 3} package.}
    \label{fig:atmosphere_corner}
\end{figure*}

The modelling results are shown in \Cref{tab:atmosphere_para}, and \Cref{fig:transmission_spectrum,fig:atmosphere_corner}. HAT-P-26~b’s atmosphere is modelled to have a 100 Pa temperature of $590^{+60}_{-50}$ K, which is cooler than the estimated equilibrium temperature ($\approx$1000 K) \citep{hartman2011}. This temperature is compatible with the calculated 100 Pa temperature of \citet{macdonald2019} ($563^{+59}_{-55}$ K). Combining the result with our cloud top pressure at $P_{\textup{cloud}} > 10^3$ Pa, HAT-P-26~b can be assumed to have a cloud- and haze-free atmosphere with the He/H$_2$ ratio of 0.1. The ratio indicates that H$_2$ dominates the atmosphere. The transmission analysis suggests the water abundance of $2.4^{+3.0}_{-1.6}$ \% H$_2$O in volume mixing ratio. While, the other modelled chemical compositions should represent less than 0.01\% in volume mixing ratio of the atmosphere.

\begin{figure*}[htb]
\centering
    \includegraphics[width=1\textwidth,page=1]{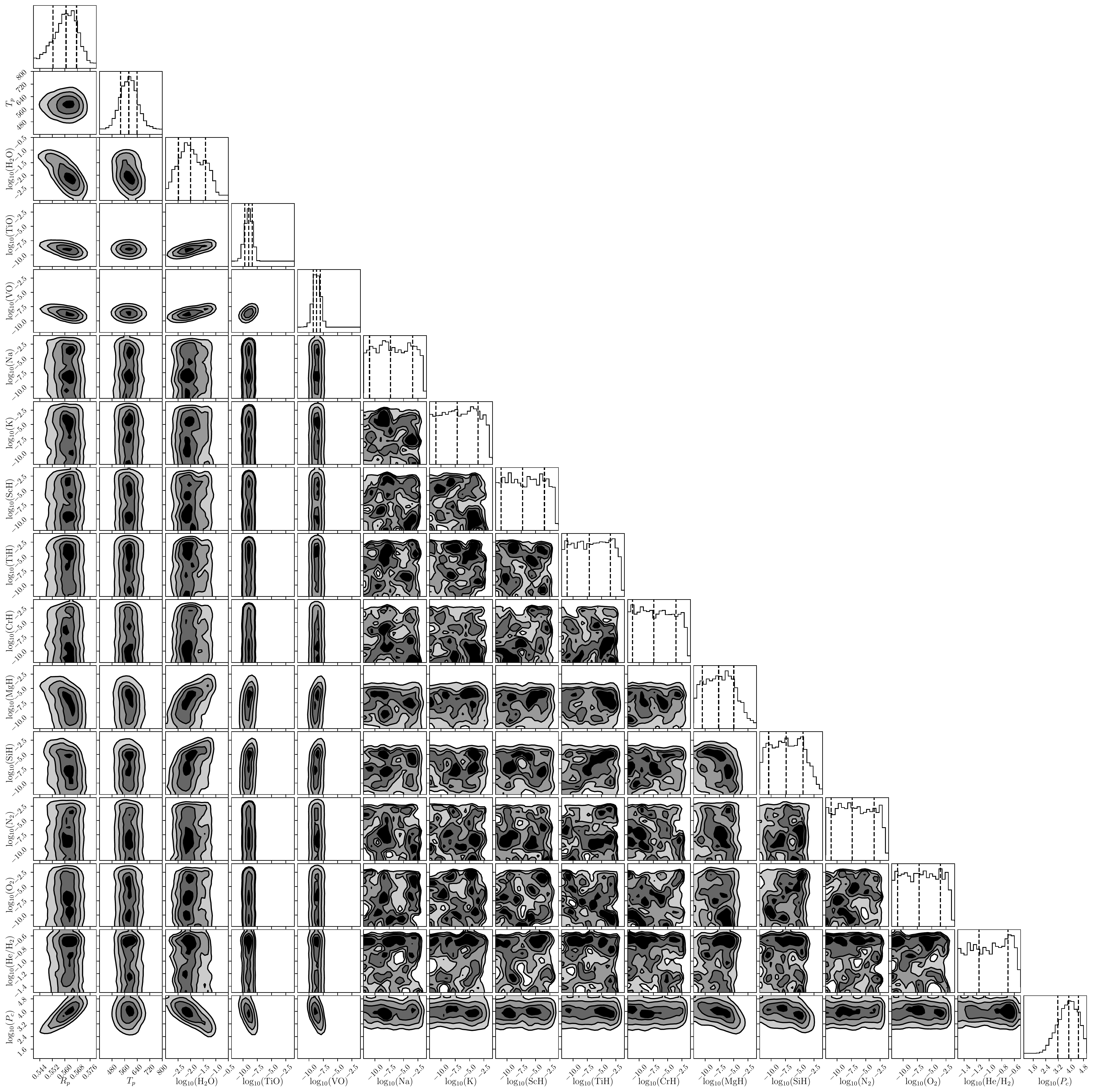} 
    \caption{Posterior probability distribution for our HAT-P-26~b atmospheric model, calculated using only the HST/WFC3 data, via nested sampling with \texttt{TauREx 3}.}
    \label{fig:atmosphere_corner_HST}
\end{figure*}

\begin{figure}[htb]
\centering
    \includegraphics[width=0.49\textwidth,page=1]{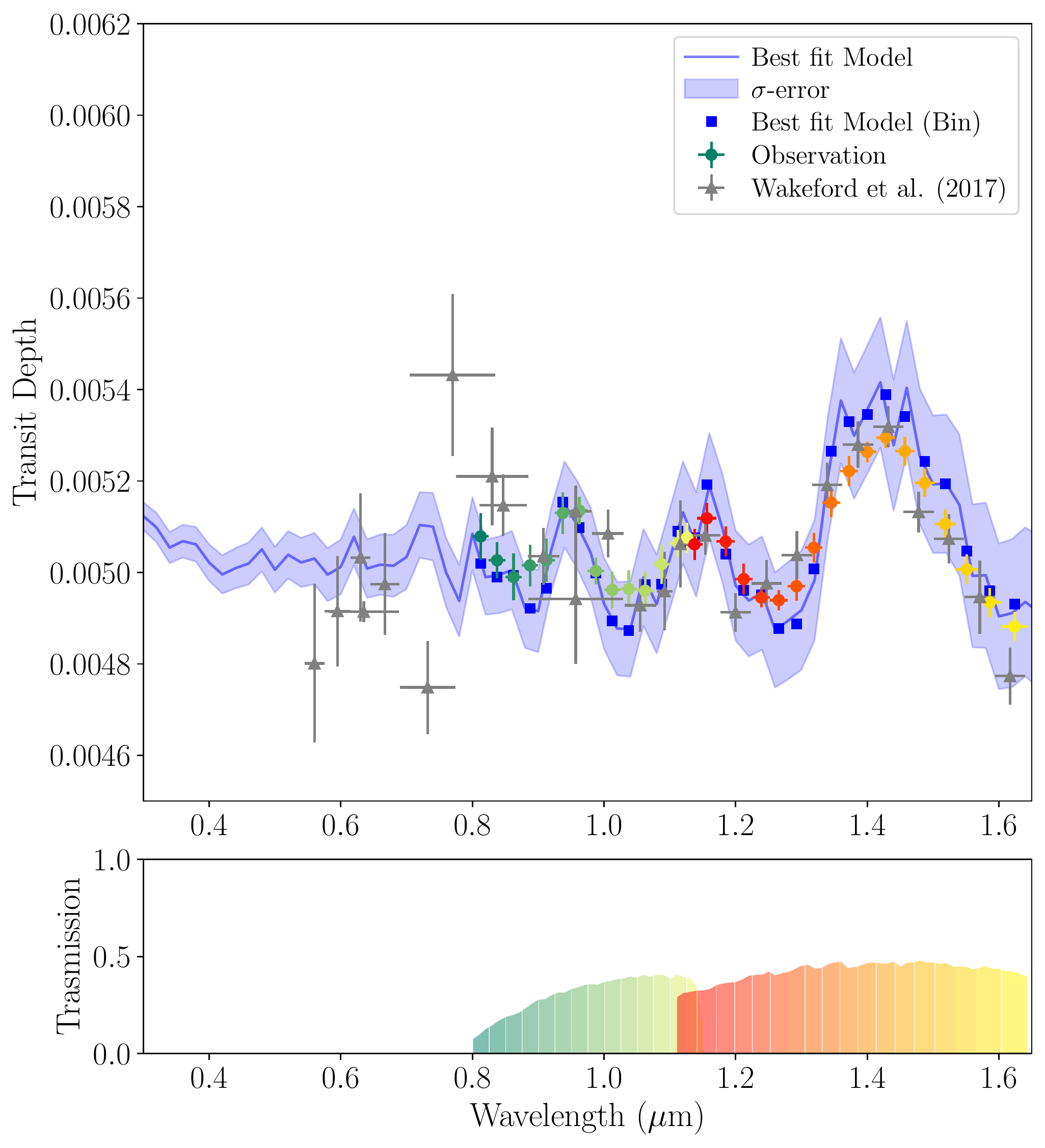} 
    \caption{Top panel: The best-fit transmission spectrum of HAT-P-26~b, based on fitting only the HST/WFC3 data from \texttt{TransitFit} with synthetic models generated by \texttt{TauREx 3} (blue solid line), with their 1$\sigma$ error (blue bands). The blue squares are the binned best-fit transmission spectra. The data are binned using the bandpass in the bottom panel.}
    \label{fig:transmission_spectrum_HST}
\end{figure}

To compare our result to \citet{macdonald2019}, which uses the same HST/WFC3 data, we employed  \texttt{TauREx 3} to model the transmission spectra exclusively from the HST/WFC3 observations. \Cref{fig:atmosphere_corner_HST,fig:transmission_spectrum_HST}, show that this model retrieves an H$_2$O abundance of $1.0^{+2.9}_{-0.6}$\% H$_2$O, which is similar to the abundance obtained by \citet{macdonald2019} ($1.5^{+2.1}_{-0.9}$ \% H$_2$O). Furthermore, both models also provide the same atmospheric temperature at 100 Pa (590 K). However, our analysis does not provide any evidence for the presence of metal hydrides, as reported by \citet{macdonald2019}.

The discrepancy observed may be attributed to the absence of HST/STIS transits and Spitzer transits in our atmospheric modeling, which were used in \citet{macdonald2019}. We have not include them as we were unable to obtain the raw light curves. Simply adding the published HST/STIS and Spitzer transit depths to our atmospheric analysis would not be a suitable solution, since those depths result from different physical parameters (orbital period, semi-major axis and inclination). However, when we modelled the transmission spectra of \citet{macdonald2019} using the \texttt{TauREx 3}, we also obtained the same chemical abundance as reported by \citet{macdonald2019} which used the POSEIDON code \citep{2017MNRAS.469.1979M} to be the retrieval model (\Cref{tab:atmosphere_para}). Therefore, the non-detection of metal hydrides in this work is not led by the difference atmospheric retrieval model.

Nevertheless, the reported detections of metal hydrides by \citet{macdonald2019} still exhibit a low abundance level (less than 0.01\% abundance), making it challenging to confirm their presence definitively. However, the retrieved abundance in our analysis remains within the 1-sigma error bar of the results obtained by \citet{macdonald2019}. Furthermore, it should be noted that we have not included the optical spectra data from the \textit{Magellan} Low Dispersion Survey Spectrograph 3 and HST STIS G750L observations, which were used in the analysis conducted by \citet{macdonald2019}. Since the strongest metal hydride features identified by \citet{macdonald2019} occur within the Magellan and HST bandpasses, our exclusion of this data may explain why we obtain non-detections. 

\begin{table*}
\begin{center}
\caption{The parameters and priors used for \texttt{TauREx 3} retrieval, and the best-fit retrieved parameters, based on fitting only the HST/WFC3 data from both \citet{macdonald2019} and \texttt{TransitFit}, and based on fitting all available datasets.}
\label{tab:atmosphere_para}          
\small\addtolength{\tabcolsep}{-2pt}
\begin{tabular}{@{\extracolsep{1pt}}lccccccc@{}}
\toprule
\multirow{3}{*}{Parameter}  & \multirow{3}{*}{Priors}  & \multirow{3}{*}{Scale} & Published Value & \multicolumn{3}{c}{Retrieved Value}   \\
\cline{4-4} \cline{5-7}
  &  &  & HST/WFC3 & HST/WFC3 & HST/WFC3 & All \\ 
  &  &  & MacDonald et al. (2019) & MacDonald et al. (2019) & \texttt{TransitFit} & \texttt{TransitFit} \\
 \hline
$R_p$ ($R_{\textup{Jup}}$)   &  (0.5, 0.6)  & linear & - & $0.570^{+0.005}_{-0.009}$ & $0.561^{+0.007}_{-0.008}$ & $0.564^{+0.005}_{-0.006}$      \\
$T$ (K)    & (400, 1200)  & linear & $563^{+59}_{-55}$ & $560^{+90}_{-60}$ & $590^{+50}_{-50}$ & $590^{+60}_{-50}$   \\
H$_{2}$O   & (-4, -0.2)   & log & $-1.83^{+0.39}_{-0.43}$ & $-1.5^{+0.4}_{-0.5}$ & $-2.0^{+0.6}_{-0.5}$ & $-1.6^{+0.3}_{-0.5}$   \\
TiO & (-12, -1)  & log  & - & $-11^{+1}_{-1}$ & $-9.0^{+0.6}_{-0.7}$ & $-10^{+1}_{-1}$   \\
VO  & (-12, -1) & log  & - & $-9^{+1}_{-2}$ & $-8.7^{+0.6}_{-0.6}$ & $-8.9^{+0.6}_{-0.7}$   \\
Na  & (-12, -1) & log  & - & $-7^{+3}_{-3}$ & $-7^{+4}_{-4}$ & $-6^{+3}_{-4}$   \\
K   & (-12, -1) & log  & - & $-7^{+3}_{-3}$ & $-7^{+4}_{-4}$ & $-7^{+4}_{-4}$   \\
ScH & (-12, -1) & log  & $-4.76^{+0.91}_{-4.09}$ & $-7^{+3}_{-3}$ & $-7^{+4}_{-4}$ & $-7^{+4}_{-3}$   \\
TiH & (-12, -1) & log  & $-6.24^{+0.71}_{-0.65}$ & $-7^{+3}_{-3}$ & $-7^{+4}_{-4}$ & $-7^{+3}_{-4}$   \\
CrH & (-12, -1) & log  & $-5.72^{+0.89}_{-1.37}$ & $-7^{+3}_{-3}$ & $-7^{+4}_{-4}$ & $-7^{+3}_{-3}$   \\
MgH & (-12, -1) & log  & - & $-8^{+3}_{-3}$ & $-8^{+3}_{-3}$ & $-6^{+3}_{-4}$   \\
SiH & (-12, -1) & log  & - & $-7^{+3}_{-3}$ & $-7^{+3}_{-3}$ & $-6^{+3}_{-4}$   \\
N$_{2}$ & (-12, -1) & log & - & $-7^{+3}_{-3}$ & $-7^{+4}_{-4}$ & $-6^{+3}_{-4}$   \\
O$_{2}$ & (-12, -1) & log & - & $-7^{+3}_{-3}$ & $-7^{+4}_{-4}$ & $-7^{+3}_{-3}$   \\
He/H$_{2}$ & (-3, -0.04) & log & $-0.77$  & $-1.6^{+0.9}_{-0.9}$ & $-1.2^{+0.5}_{-0.5}$ & $-1.0^{+0.3}_{-0.4}$   \\
$P_{\textup{clouds}}$ (Pa) & (1, 5)  & log  & $>0.72$ & $3.9^{+0.6}_{-0.8}$ & $3.9^{+0.6}_{-0.7}$ & $4.2^{+0.5}_{-0.6}$
\\
\hline
\end{tabular}
\end{center}
\end{table*}

\section{Discussion and Conclusions}
\label{sec:conclusion}
This work performs multi-band photometric follow-up observations of the Neptune-mass planet HAT-P-26 b, using a range of space and ground-based data, including new data gathered from the SPEARNET telescopes network. A total of 13 new transit light curves were combined with published light curves from HST, TESS, and ground-based telescopes, to model the physical parameters of HAT-P-26~b using the \texttt{TransitFit} light-curve analysis package. 

By fitting these observations, we derived the following parameters of HAT-P-26 b: an inclination of $i = 87.83 \pm 0.05$ deg, a star--planet separation of 12.49 $\pm$ 0.07 $R_\ast$, plus the mid-transit times for each transit event and the planet-to-star radius ratio ($R_p$/$R_\ast$) for each filter. Limb-darkening parameters for the HST/WFC3 G102 and G104 grism data are compatible with the computed values from the ExoCTK. However, the fitted optical limb-darkening from \texttt{TransitFit} shows inconsistency with the ExoCTK calculated values.

Based on the mid-transit times from 33 epochs obtained from \texttt{TransitFit}, we refined the linear ephemeris, finding $T^{c}_{m}(E) = 2455304.65211^{+0.00036}_{-0.00035} + E\; 4.234503^{+0.000001}_{-0.000001} $. We performed a periodogram analysis to search for TTV signals that might be caused by an additional planet in the HAT-P-26 system. A TTV amplitude of 1.98 $\pm$ 0.05 minutes was detected with a frequency of 0.0045 $\pm$ 0.0001 cycle/period, equivalent to a sinusoidal period of $\simeq222$ epoch. This is shorter than the period presented by \citet{von2019} ($\simeq270$ epoch). If the TTV amplitude is due to the presence of a third-body orbit that is near the first-order resonance of HAT-P-26 b ($\sim$8.47 days), its mass could be around 0.02M$_{\textup{Jup}}$ (6.36M$_{\oplus}$). 

The atmospheric composition of HAT-P-26b is modeled using the transit depths obtained from the \texttt{TransitFit} package and analyzed with \texttt{TauREx3}. At a pressure of 100 Pa, HAT-P-26b exhibits an atmospheric temperature of $590^{+60}_{-50}$ K, with a cloud-top pressure estimated to be $P_c > 10^4$ Pa. The abundance of H$_2$O in HAT-P-26b's atmosphere is determined to be $2.4^{+3.0}_{-0.6}$\% in the volume mixing ratio, which aligns with the abundance reported by \citet{macdonald2019}. Although other modeled chemical components are expected to contribute less than 0.01\% in the volume mixing ratio to the overall atmosphere and do not indicate clear evidence in support of the presence of metal hydrides as reported by \citet{macdonald2019}), our analysis yields an abundance within the 1-sigma error range of \citet{macdonald2019}. This discrepancy in the detection is not attributed to the difference in the atmospheric retrieval model used. Nevertheless, the absence of detected metal hydrides in our study could still be attributed to differences in the optical spectra used for the analysis in our work and in the study by \citet{macdonald2019}. 

\section*{Acknowledgments}

\begin{acknowledgments}
We are grateful to the anonymous referee for useful suggestions, which help to improve this paper significantly. 
This work is based on the observations made with ULTRASPEC at the Thai National Observatory and the Thai Robotic Telescopes operated by the National Astronomical Research Institute of Thailand (Public Organization). The data used in this work also included the available data based on observations with the NASA/ESA Hubble Space Telescope, obtained at the Space Telescope Science Institute (STScI) operated by AURA, Inc. The publicly available HST observations presented here were taken as part of proposal 14260, led by PI: Drake Deming and proposal 14110, led by David Sing. These were obtained from the Hubble Archive. Additionally, this work included the data collected by the TESS mission, which the funding provided by the NASA Explorer Program. HST and TESS data presented in this paper were obtained from the Mikulski Archive for Space Telescopes (MAST) at the Space Telescope Science Institute. The specific observations analyzed can be accessed via \dataset[https://doi.org/10.17909/pmhe-db52]{https://doi.org/10.17909/pmhe-db52} and \dataset[https://doi.org/10.17909/t9-nmc8-f686]{https://doi.org/10.17909/t9-nmc8-f686}. This research made use of the open source Python package exoctk, the Exoplanet Characterization Toolkit \citep{ExoCTK}.

This work presents results from the European Space Agency (ESA) space mission \textit{Gaia}. \textit{Gaia} data are being processed by the \textit{Gaia} Data Processing and Analysis Consortium (DPAC). Funding for the DPAC is provided by national institutions, in particular the institutions participating in the \textit{Gaia} MultiLateral Agreement (MLA). The \textit{Gaia} mission website is \texttt{https://www.cosmos.esa.int/gaia}. The \textit{Gaia} archive website is \texttt{https://archives.esac.esa.int/gaia}. We thank Angelos Tsiaras and Quentin Changeat for the suggestion on \texttt{Iraclis} and the instruction of \texttt{TauREx}.

This work is supported by the grant from the Ministry of Science and Technology (MOST), Taiwan. The grant numbers are MOST 109-2112-M-007-007, MOST 110-2112-M-007-035, and MOST 111-2112-M-007-035. This work is also partially supported by National Astronomical Research Institute of Thailand (Public Organization) research grant. 
\end{acknowledgments}
\vspace{5mm}
\facilities{HST/WFC3 (G141 and G102), TESS, 2.4-m (TNT), 0.5-m (TRT-TNO), 0.7-m (TRT-GAO) and 0.7-m (TRT-SRO)}


\software{\texttt{sextractor} \citep{berlin1996}, \texttt{Astrometry.net} \citep{lang2010}, \texttt{TransitFit} \citep{hay2021}, \texttt{Iraclis} \citep{tsiaras2016} and \texttt{TauREx} \citep{al-rafaie2021}.}


\appendix
\counterwithin{figure}{section}
\section{Individual SPEARNET transit light curves.}
Individual SPEARNET transit light curves of HAT-P-26 b from the observations in 2015-2018 are presented here.

\begin{figure*}[htb]
\centering
  \begin{tabular}{cc}
    \includegraphics[width=0.7\textwidth,page=1]{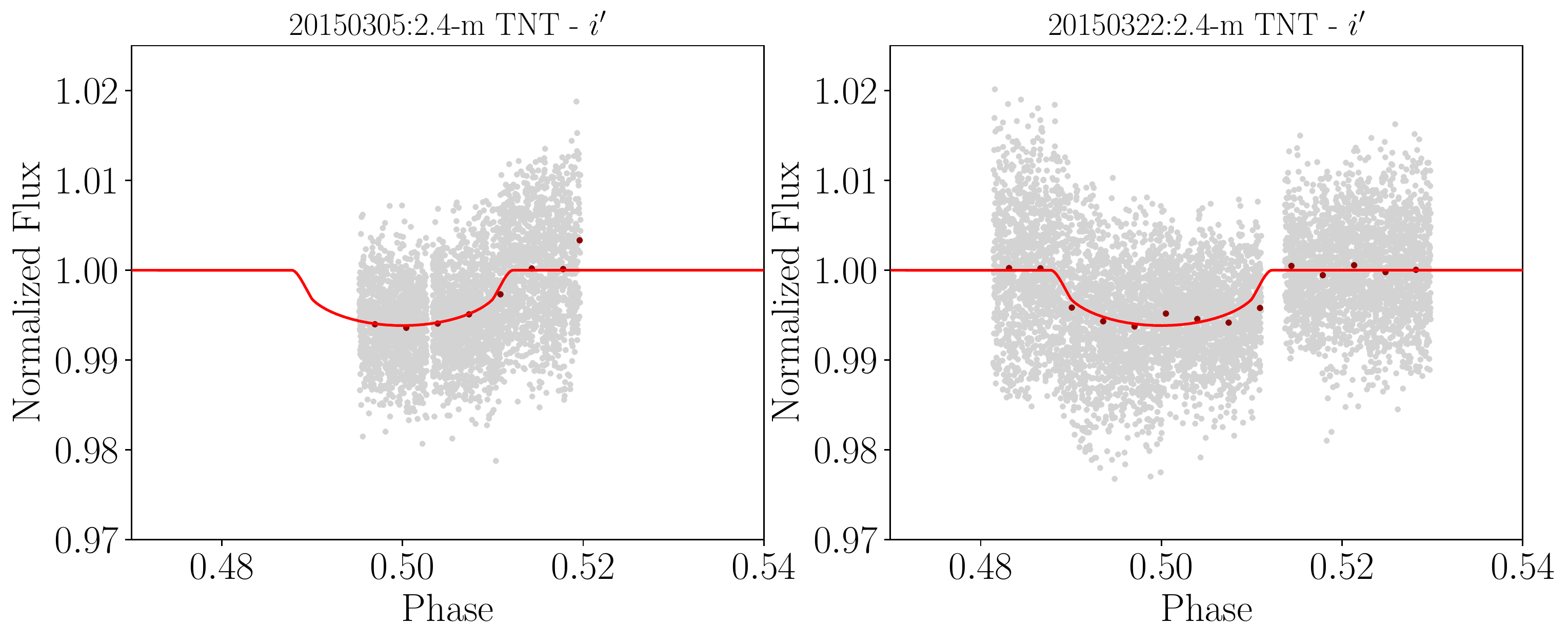} \\
    \includegraphics[width=0.7\textwidth,page=1]{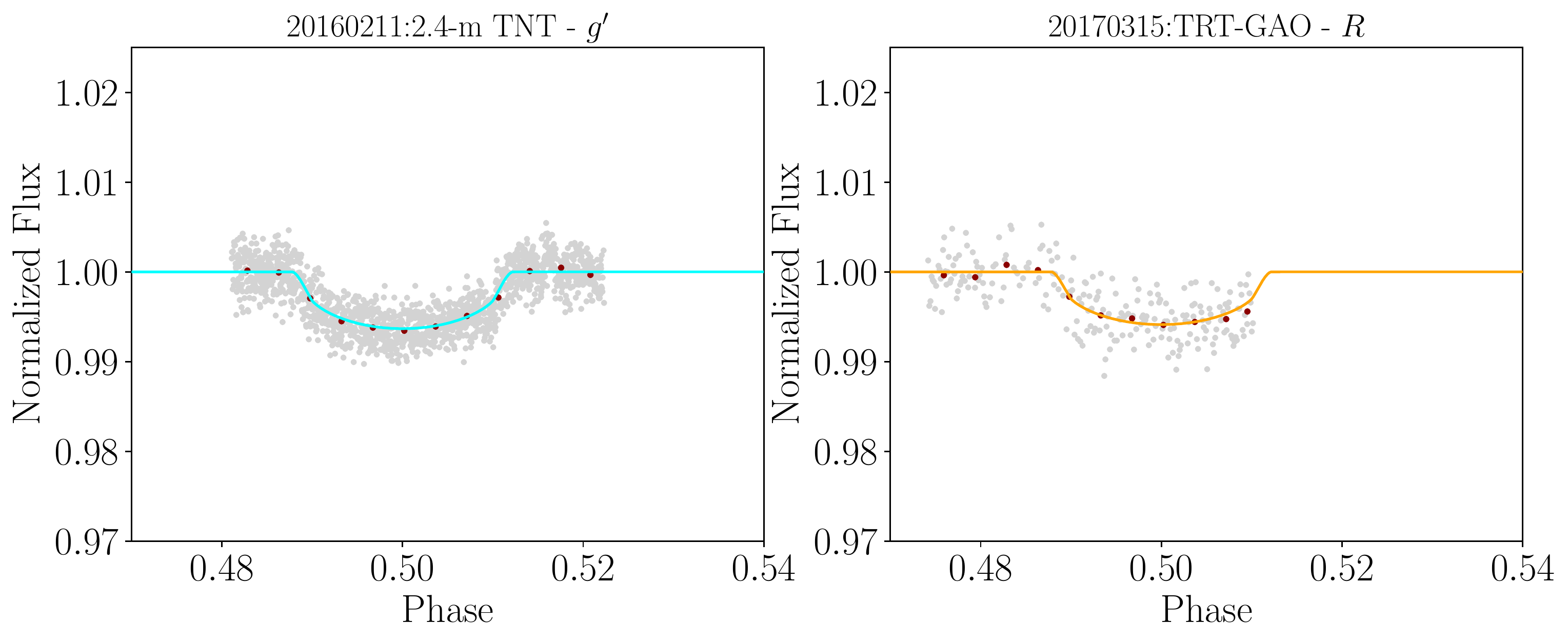} \\
    \includegraphics[width=0.7\textwidth,page=1]{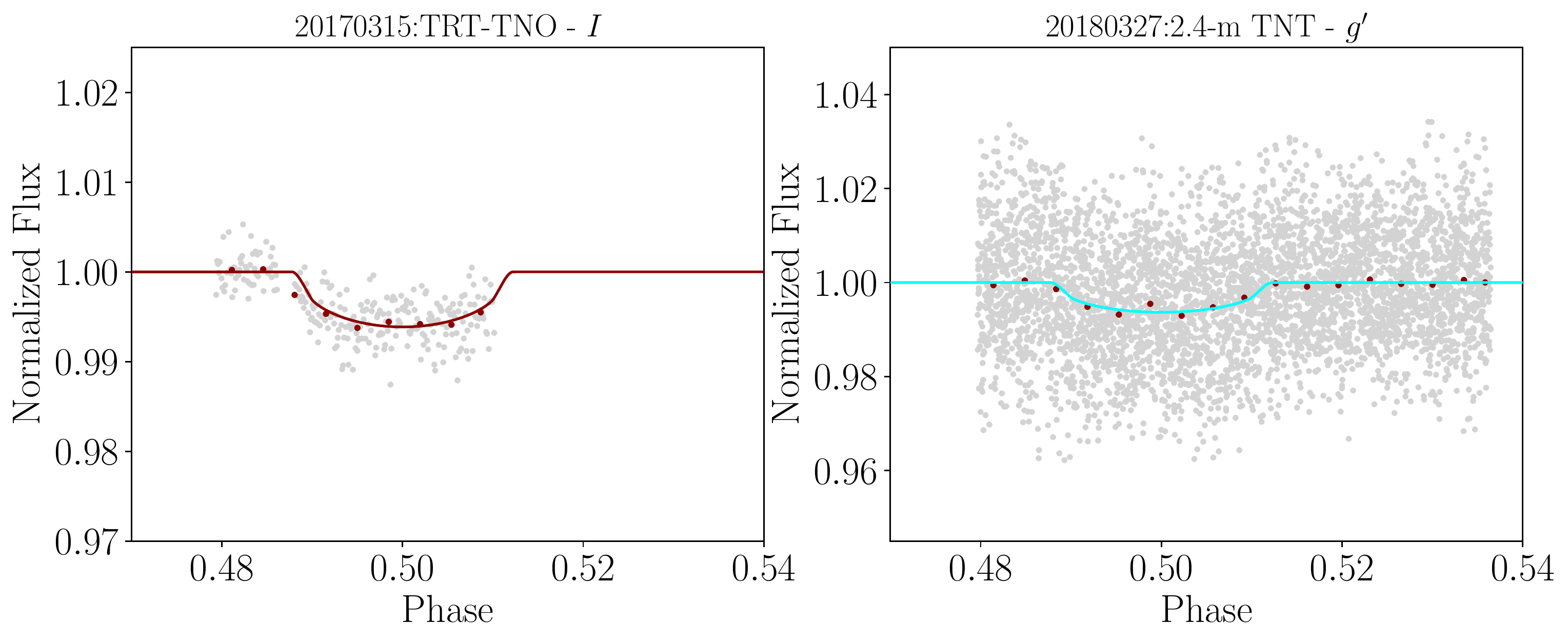} \\
    \includegraphics[width=0.7\textwidth,page=1]{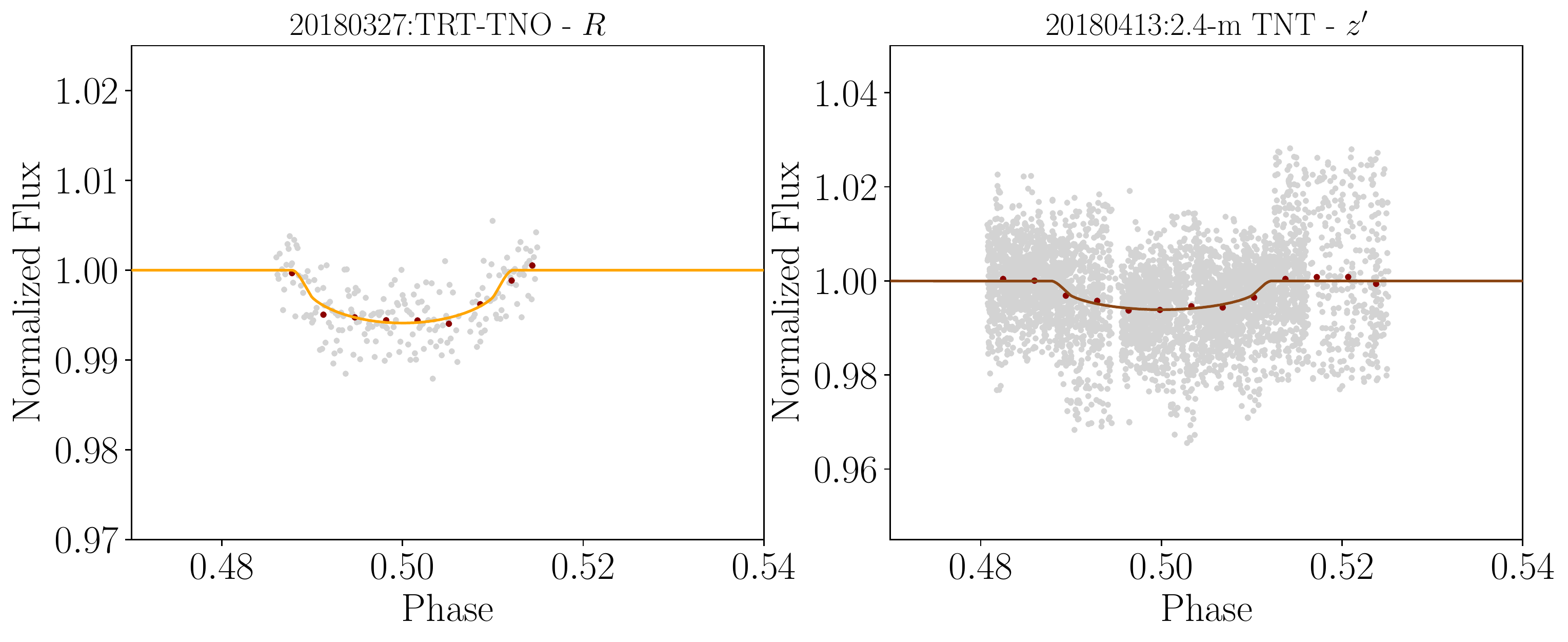} &
       \end{tabular}
    \caption{Individual SPEARNET transit light curves of HAT-P-26 b from the observations in 2015-2018. The light curves are normalized (Gray dot) and modeled by \texttt{TransitFit} (Solid line). The light curves are observed in $g'$ (Blue), $r'$ (Yellow), $R$ (Orange), $i'$ (Red), $I$ (Dark-red), and $z'$ (Brown) filters. The 5-min binned light curves are shown in the red dot.}
    \label{fig:LCs_TNT-Individuals-2015-2018}
\end{figure*}

\begin{figure*}[htb]
\centering
  \begin{tabular}{cc}
    \includegraphics[width=0.7\textwidth,page=1]{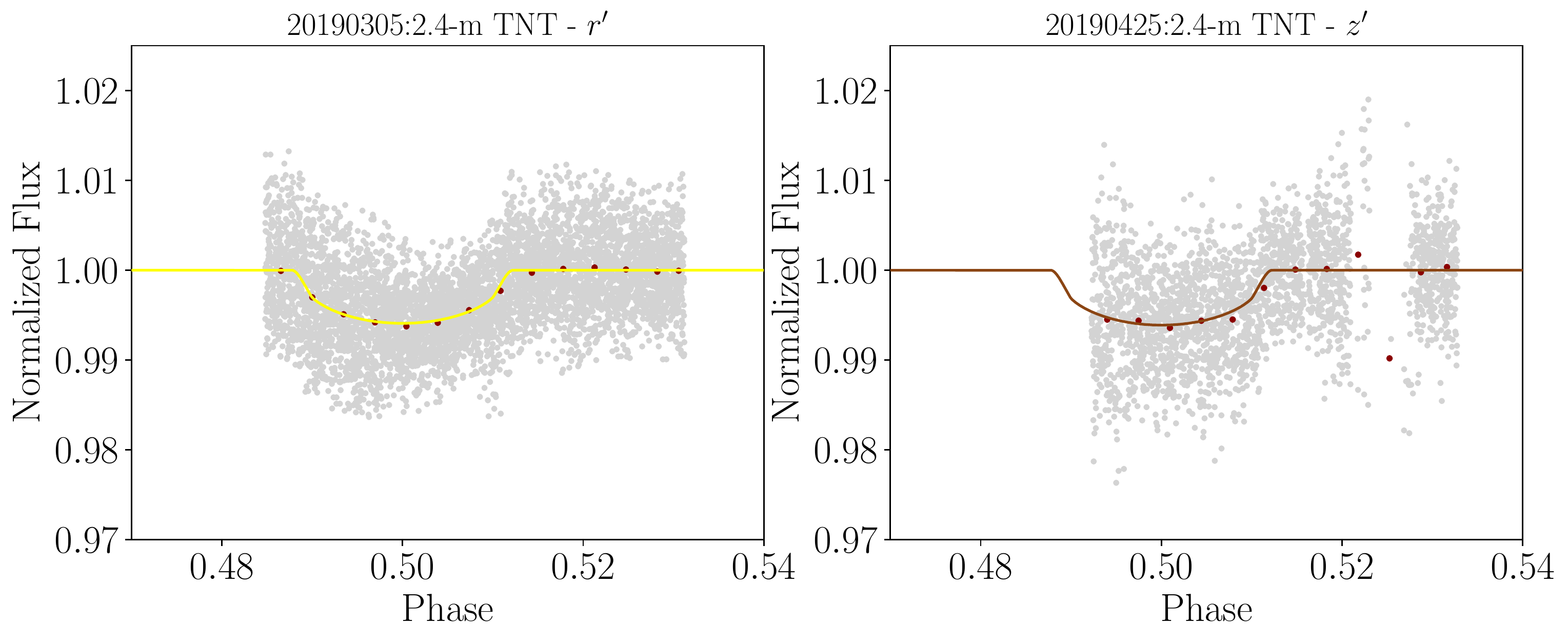} \\
    \includegraphics[width=0.7\textwidth,page=1]{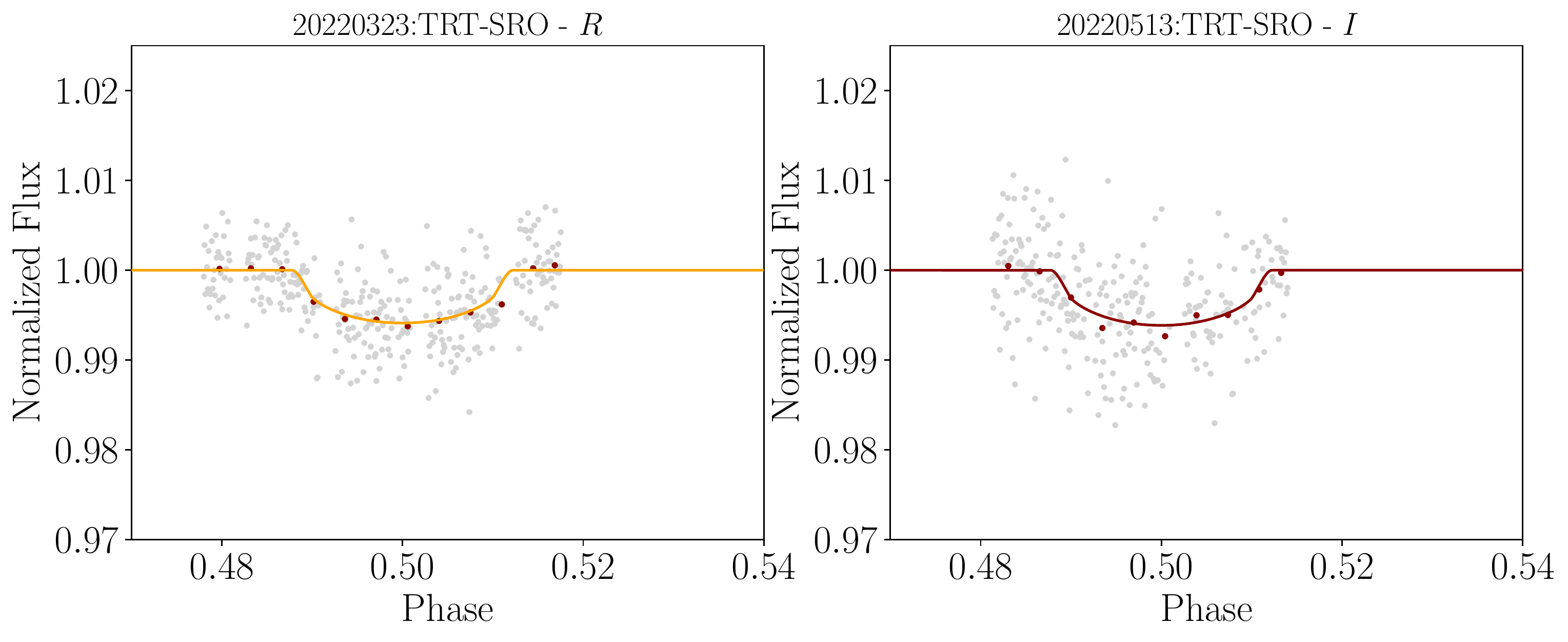} \\
    \includegraphics[width=0.35\textwidth,page=1]{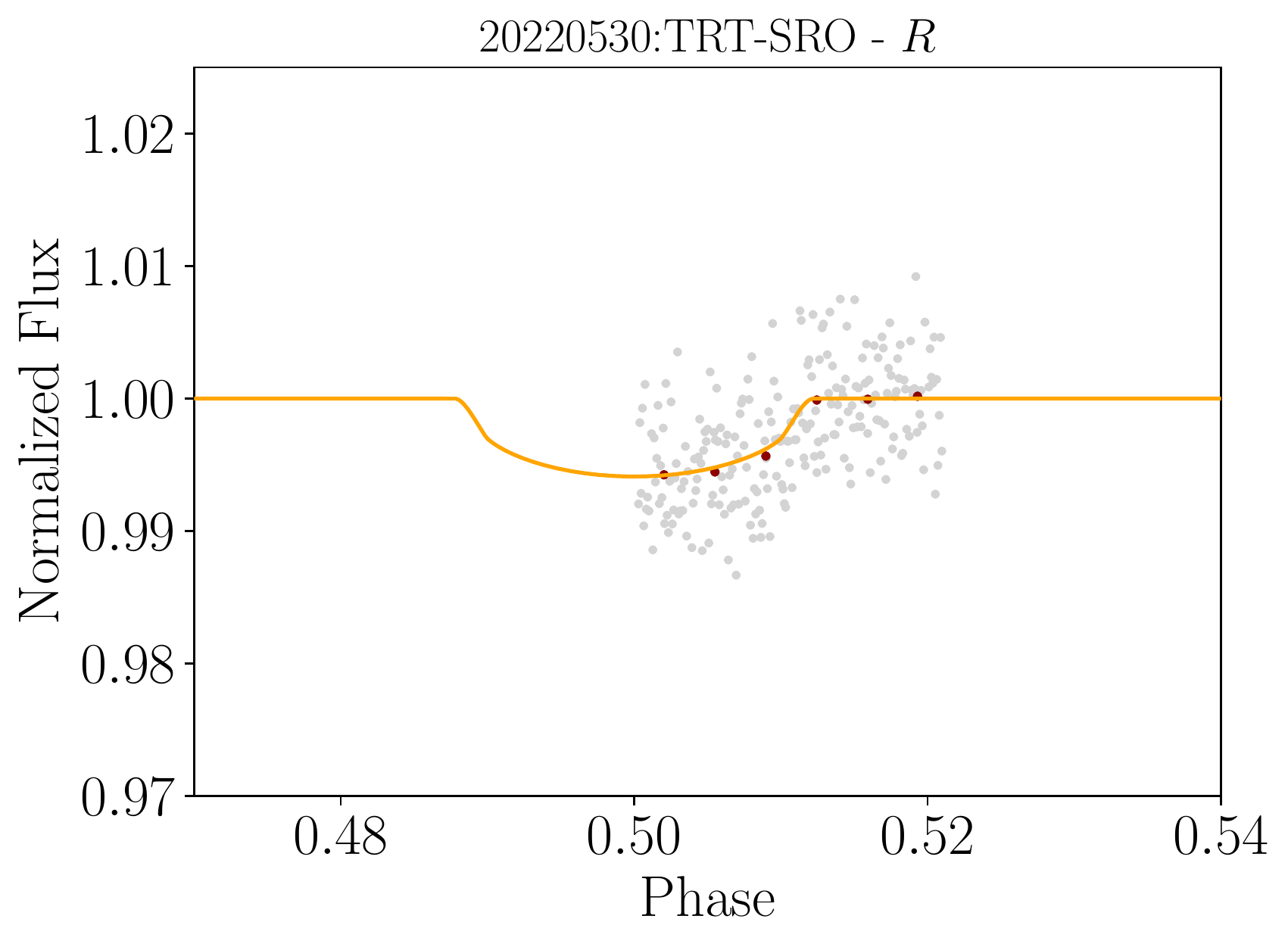} &
    \end{tabular}
    \raggedright
     \caption{Individual SPEARNET transit light curves of HAT-P-26 b from the observations in 2015-2018. The light curves are normalized (Gray dot) and modeled by \texttt{TransitFit} (Solid line). The light curves are observed in $g'$ (Blue), $r'$ (Yellow), $R$ (Orange), $i'$ (Red), $I$ (Dark-red), and $z'$ (Brown) filters. The 5-min binned light curves are shown in the red dot. (Continue)}
\end{figure*}

\section{Posterior probability distribution for the linear ephemeris model MCMC fitting parameters.}

Here we present posterior probability distributions of the linear ephemeris MCMC fitting parameters.

\begin{figure*}[h]
\begin{center}
\includegraphics[width=0.49\textwidth]{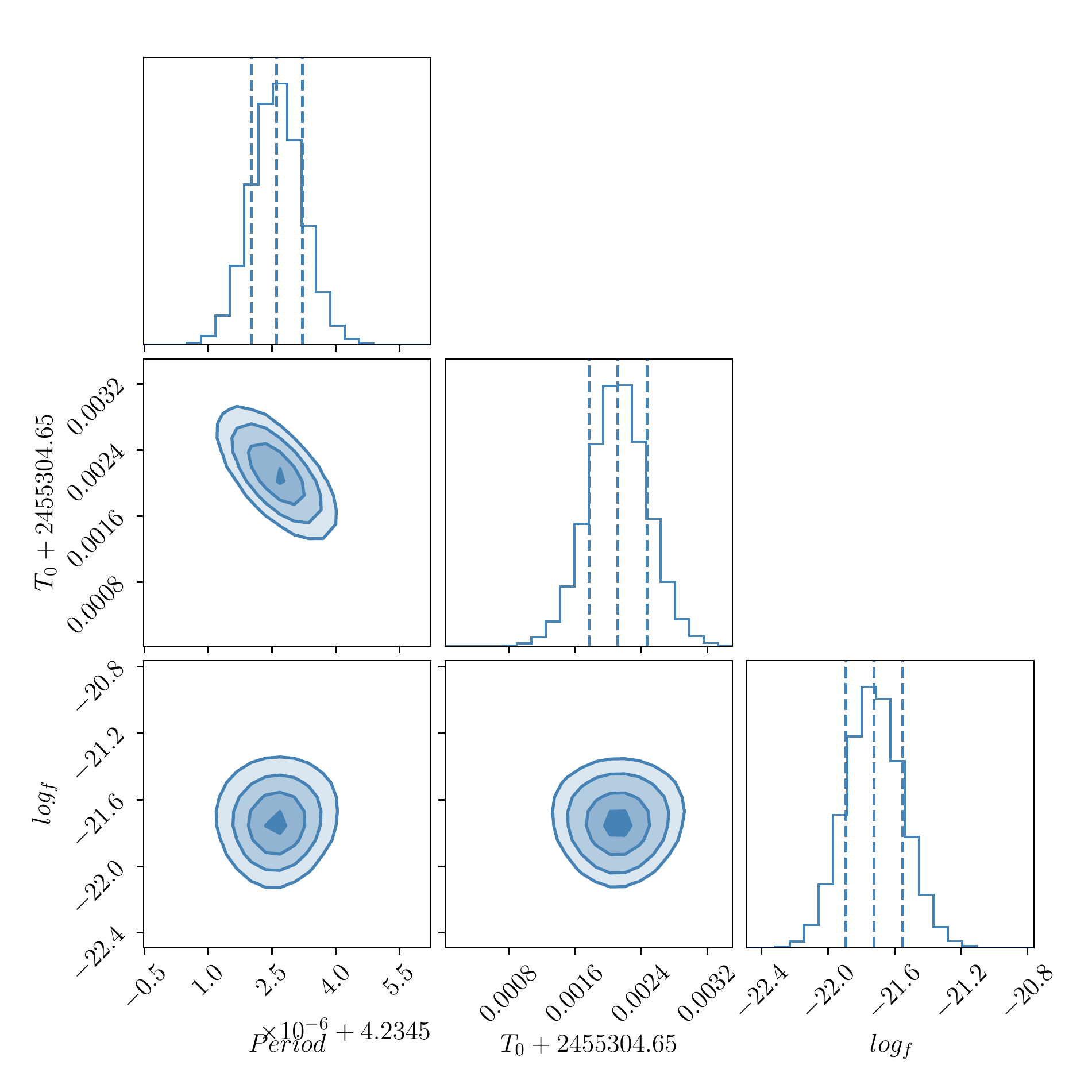}
\caption{Posterior probability distribution of the linear ephemeris MCMC fitting parameters for the 33 mid-transit times obtained from the \texttt{TransitFit}.}
\label{fig:linearMCMC_TransitFit}
\end{center}
\end{figure*}
\begin{figure*}[h]
\begin{center}
\includegraphics[width=0.49\textwidth]{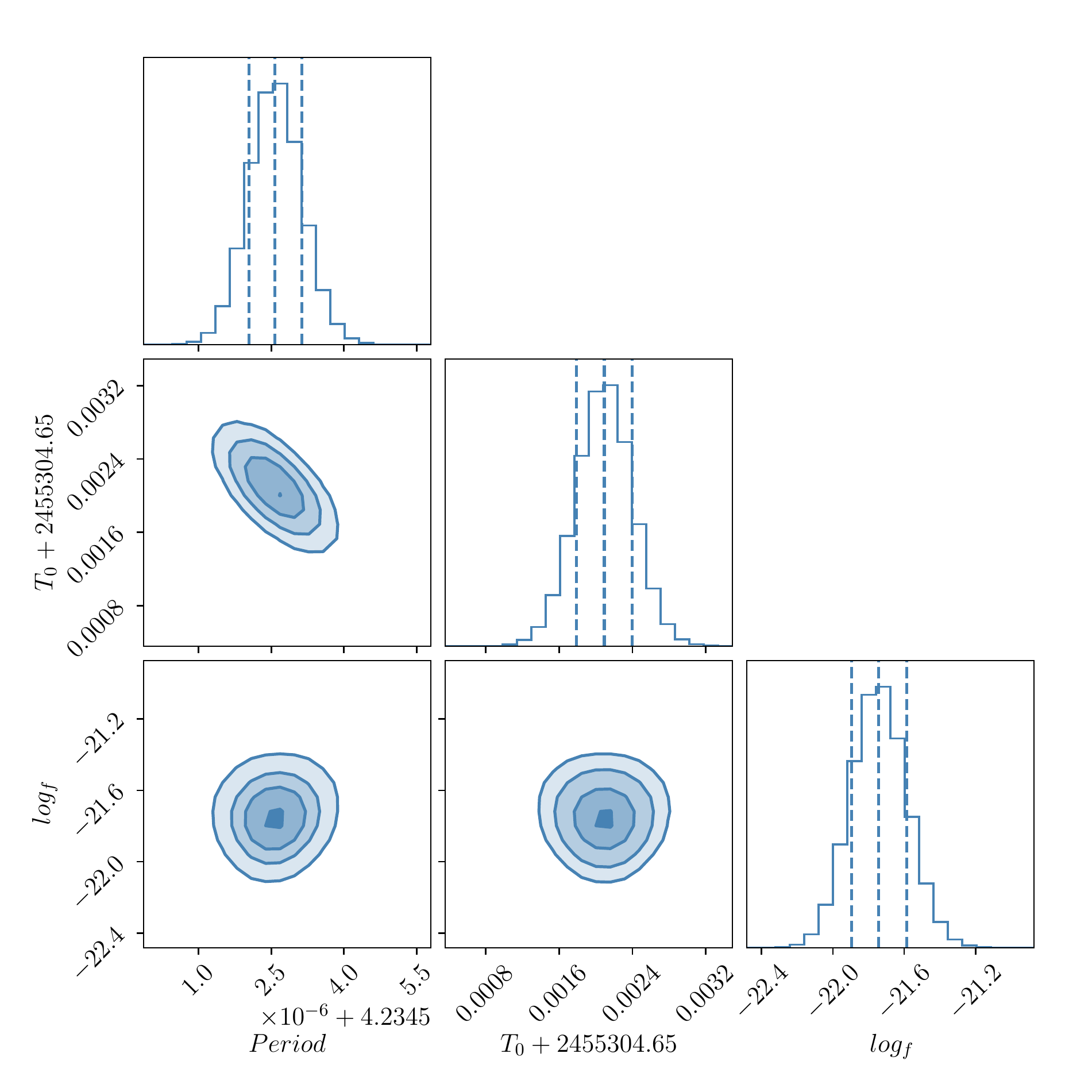}
\caption{Posterior probability distribution of the linear ephemeris MCMC fitting parameters for considering all 39 mid-transit times.}
\label{fig:linearMCMC_TransitFitLit}
\end{center}
\end{figure*}



\bibliography{HATP26b}{}
\bibliographystyle{aasjournal}



\end{document}